\documentclass[twocolumn]{aastex63}
\usepackage[utf8]{inputenc}
\usepackage{graphics,graphicx}
\usepackage{multirow}
\usepackage{bm}
\usepackage{newtxtext,newtxmath}
\usepackage{amsmath}
\usepackage{epsfig}
\usepackage{url}
\usepackage{wrapfig}
\usepackage{float}
\usepackage{subfigure}
\usepackage{xcolor}
\usepackage{longtable}

\defcitealias{Hasan20}{H20}
\defcitealias{Davies23b}{D23}
\defcitealias{Li24}{HIERACHY II}

\shortauthors{X.-D. Yu et al.}
\shorttitle{HIERACHY III: CIV catlog}

\begin{document}

\title{Probing the \ion{He}{2} reionization ERa via Absorbing \ion{C}{4} Historical Yield (HIERACHY) III: The \ion{C}{4} absorber catalog and initial results on cosmic abundance evolution at $\bm{z\approx 3-5}$}

\author{Xiaodi Yu}
\affiliation{Department of Astronomy, Tsinghua University, Beijing 100084, People’s Republic of China}
\affiliation{Purple Mountain Observatory, Chinese Academy of Sciences, 10 Yuanhua Road, Nanjing 210023, People’s Republic of China}

\author[0000-0002-2941-646X]{Zhijie Qu}
\affiliation{Department of Astronomy \& Astrophysics, The University of Chicago, 5640 S. Ellis Ave., Chicago, IL 60637, USA}
\affiliation{Department of Astronomy, Tsinghua University, Beijing 100084, People’s Republic of China}

\author{Zheng Cai}\thanks{E-mail: zcai@mail.tsinghua.edu.cn (CZ)}
\affiliation{Department of Astronomy, Tsinghua University, Beijing 100084, People’s Republic of China}

\author[0000-0001-6239-3821]{Jiang-Tao Li}\thanks{E-mail: pandataotao@gmail.com (LJT)}
\affiliation{Purple Mountain Observatory, Chinese Academy of Sciences, 10 Yuanhua Road, Nanjing 210023, People’s Republic of China}

\author[0009-0006-7138-2095]{Huiyang Mao}
\affiliation{Purple Mountain Observatory, Chinese Academy of Sciences, 10 Yuanhua Road, Nanjing 210023, People’s Republic of China}

\author[0000-0002-9373-3865]{Xin Wang}
\affil{School of Astronomy and Space Science, University of Chinese Academy of Sciences (UCAS), Beijing 100049, People’s Republic of China}
\affil{National Astronomical Observatories, Chinese Academy of Sciences, Beijing 100101, People’s Republic of China}
\affil{Institute for Frontiers in Astronomy and Astrophysics, Beijing Normal University,  Beijing 102206, People’s Republic of China}

\begin{abstract}
As part of the HIERACHY program, we collect the high-SN and high-spectral resolution optical spectra of 25 quasars at $z\approx4-5$ to constrain the \ion{C}{4} evolution at $z\approx 3-5$.
In this paper, we report a catalog of 626 (1263) \ion{C}{4} absorption systems (components) at $z\approx3-5$ with a 50\% completeness column density of log$(N_{\rm CIV}/\rm cm^{-2}) \approx 12.3$. 
The HIERACHY/MIKE \ion{C}{4} sample is one of the best \ion{C}{4} absorber samples optimized to study the IGM during the \ion{He}{2} reionization epoch.
Using 557 (1090) intervening absorption systems (components), we found the column density distribution function of \ion{C}{4} absorption systems with log$(N_{\rm CIV}/\rm cm^{-2})\gtrsim 12.3$ has a broken power-law shape, with the turn-over column density log$(N_{\rm crit}/\rm cm^{-2}) = 13.35^{+0.20}_{-0.19}$, which is close to or smaller than the detection limit of most literature samples.
We also found that both comoving path length number density $dn/dX$ and cosmic abundance $\Omega$ for \ion{C}{4} absorption systems with log$(N_{\rm CIV}/\rm cm^{-2})> 13.2$ show an increase (at the 2.2$\sigma$ and 1.4$\sigma$ levels, respectively) from redshift $z\approx5$ to 3,
while absorption systems with log$(N_{\rm CIV}/\rm cm^{-2})= 12.3-13.2$ exhibit a constant $dn/dX$ and $\Omega_{\rm CIV}$.
\end{abstract}


\section{Introduction}

\ion{He}{2} reionization has been proposed to be another major heating event for the bulk of baryonic matter in the Universe (e.g., \citealt{Miralda00,McQuinn09,McQuinn16}), following the end of \ion{H}{1} reionization at $z\approx6$ (e.g., \citealt{Fan06,Jiang22}). The existence of \ion{He}{2} reionization ending at $z\approx3$ is supported by the observations of \ion{He}{2} Ly$\alpha$ ($\lambda_{\rm rest} = 303.78$~\AA) optical depth  (e.g., \citealt{Worseck19,Villasenor22}) and the measurements of thermal state for IGM \citep{Ricotti00,Schaye00,McDonald01,Zaldarriaga01,Theuns02,Becker11,Boera14,Hiss18,Walther19,Upton20,Gaikwad21,Villasenor22,Yang23}. 
However, the constraints on \ion{He}{2} reionization history remain limited by both the sample size of \ion{He}{2} Ly$\alpha$ absorption (\citealt{Worseck19}) and the scatter in the measured thermal state of the IGM (e.g., \citealt{Becker11,Hiss18,Walther19,Upton20}). 
As an alternative, metal absorption lines can provide additional constraints as their properties are linked to the shape and intensity of the ionizing UV background (UVB, e.g., \citealt{Bergeron86,Agafonova05,Finlator15}).

The \ion{C}{4} $\lambda\lambda 1548, 1550$ doublet is one of the most suitable metal tracers during the epoch of \ion{He}{2} reionization (\ion{He}{2} EoR, e.g., \citealt{Huscher24}). 
First, the ionization potential of the $\rm C^{3+}$ ion (64.5~eV) is comparable to that of $\rm He^+$ (54.5~eV).
Second, 
the ground-based optical spectral surveys could cover the \ion{C}{4} absorption signals ($\approx 6200-9300$~\AA~in observed frame) at the redshift of \ion{He}{2} EoR (e.g., $z\approx3-5$, \citealt{McQuinn09, Compostella14, Kulkarni19}).
Futhermore, absorption line surveys for \ion{C}{4} generally have high survey efficiency due to the relatively large redshift path that is free from the contamination of complex Ly$\alpha$ forests.
Lastly, the doublet absorption feature ensures reliable detection, even when there is significant contamination in the absorption signals.

Current \ion{C}{4} absorption surveys have covered the redshift up to 
$z\lesssim 6.5$ (\citealt{Sargent88, Chen01, Songaila01,Boksenberg03, Becker09,Ryan-Weber09,Cooksey10,Simcoe11,Cooksey13,D'Odorico13,D'Odorico2022,Boksenberg15, Burchett15,Diaz16,Cooper19,Meyer19}; \citealt[][hereafter \citetalias{Hasan20}]{Hasan20}; \citealt[][hereafter \citetalias{Davies23b}]{Davies23b}), which indicate that the cosmic abundance of \ion{C}{4} increases as redshift decreases.
However, the redshift survey path of these studies generally peaks outside $z\approx3-5$ (e.g., \citealt{D'Odorico10}; \citetalias{Hasan20,Davies23b}), leaving a significant gap in the number of observed \ion{C}{4} absorbers at $z\approx4$. This observational limitation hinders our understanding of \ion{C}{4} absorber evolution during the \ion{He}{2} EoR.

The ``Probing the \ion{\textbf{H}e}{2} re-\textbf{I}onization \textbf{ER}a via \textbf{A}bsorbing \ion{\textbf{C}}{4} \textbf{H}istorical \textbf{Y}ield (HIERACHY)'' program is designed to study the \ion{He}{2} EoR and relevant physical processes at $z\approx3-5$. 
The summary of the program is presented in \citep[][hereafter \citetalias{Li24}]{Li24}, 
including the scientific motivation, overall design, current status, major goals, and examples of initial data products. 
We herein investigate the evolution of cosmic \ion{C}{4} abundance (log$(N_{\rm CIV}/\rm cm^{-2})\gtrsim12.3$) over \ion{He}{2} EoR and present a catalog of \ion{C}{4} absorption systems using the high-resolution Magellan/MIKE spectra of 25 quasars at $z\approx3.9-5.2$. The measured cosmic \ion{C}{4} abundance evolution reflects the interplay of several astrophysical processes, including the expansion of the Universe, metal enrichment, and ionization by UVB or local ionizing sources \citep[e.g.,][]{Finlator15}. Further studies on the primary driver behind the observed evolution trends will be presented in the following papers.

The paper is structured as follows.
We describe observations and data processing procedures in \S\ref{sec:dat_ana}. 
We compare our \ion{C}{4} absorption line sample to other samples at $z\approx 3-5$ in \S\ref{sec:sample}.
In \S\ref{sec:stat}, we report the measured column density distribution function (CDDF), the comoving path length number density, and the cosmic mass abundance.
Detailed descriptions of the \ion{C}{4} absorber catalog are provided in \S\ref{sec:catalog}, and we summarize our main results in \S\ref{sec:Summary}.
Throughout the paper, we adopt a cosmological model with $H_{\rm 0}=67.7\rm~km~s^{-1}~Mpc^{-1}$, $\Omega_{\rm M}=0.3$ and $\Omega_{\rm \Lambda}=0.7$. All uncertainties are quoted at a confidence level of $1~\sigma$ unless otherwise noted.

\section{Data reduction and analysis} \label{sec:dat_ana}

In this paper, we present the \ion{C}{4} absorber sample using the high-resolution Magellan/MIKE spectra (the Magellan Inamori Kyocera Echelle spectrograph) of 25 quasars distributed in a redshift range of $z\approx3.9-5.2$.
As introduced in \citetalias{Li24}, we observed 26 quasars with Magellan/MIKE spectroscopy.
In this paper, we exclude J094604+183539 due to the poor data quality.
In a future paper, we will present additional medium-resolution spectra taken with Magellan/MagE (Magellan Echellette spectrograph).
Here, we summarize the inspection of the identified \ion{C}{4} absorber and the following analyses, while the sample selection, observation log, data quality, data reduction, continuum fitting, initial search, and identification of various absorption lines, as well as some examples of the identified lines can be found in \citetalias{Li24}.

\begin{table*}
\centering
\begin{tabular}{cccc}
\hline\hline
\multicolumn{2}{c}{\textbf{Confidence Level}} & \textbf{Descriptions} & \textbf{Criteria}\\ 
\hline
\multirow{4}{*}{$\rm Lv_{\rm fit}$} & -1 & low significance (non-detection) & \parbox{3.5in}{\centering $N_{\rm CIV}/\Delta N_{\rm CIV}<3$ in fit} \\ \cline{2-4}
 & 1 & high-level contamination & \parbox{3.5in}{\centering greater than half number of pixels within $|v| < 0.5v_{\text{FWHM}}$ of modeled line profile are $\rm {C_{p}}$}  \\ \cline{2-4}
 & 2 & low-level contamination & \parbox{3.5in}{\centering $\rm {C_{p}}$ are within the $|v| < v_{\text{FWHM}}$ of modeled \ion{C}{4} profile} \\ \cline{2-4}
 & 3 & no contamination & \parbox{3.5in}{\centering no $\rm {C_{p}}$ within $|v| < v_{\text{FWHM}}$ of modeled \ion{C}{4} line profile} \\ 
 \hline
\multirow{5}{*}{$\rm Lv_{\rm MCMC}$} & -1 & false-positive & \parbox{3.5in}{\centering other ion absorption or mismatched profiles} \\ \cline{2-4}
 & 0 & low significance & \parbox{3.5in}{\centering $N_{\rm CIV}/\Delta N_{\rm CIV}<3$ in MCMC} \\ \cline{2-4}
 & 1 & unknown contamination sources & \parbox{3.5in}{\centering most from $\rm Lv_{\rm fit} = 1$\\ $\rm {C_{p,tel}}$, $\rm {C_{p,oth-sky}}$, $\rm {C_{p,oth-CIV}}$ and $\rm {C_{p,oth-ion}}$ could not explain the contamination} \\ \cline{2-4}
 & 1.5 & known contamination sources & \parbox{3.5in}{\centering most from $\rm Lv_{\rm fit} = 1$\\ at least one of $\rm {C_{p,tel}}$, $\rm {C_{p,oth-sky}}$, $\rm {C_{p,oth-CIV}}$ and $\rm {C_{p,oth-ion}}$ could explain the contamination} \\ \cline{2-4}
 & 2 & low-level contamination & \parbox{3.5in}{\centering similar to $\rm Lv_{\rm fit} = 2$}\\ \cline{2-4}
 & 3 & no contamination & \parbox{3.5in}{\centering similar to $\rm Lv_{\rm fit} = 3$}\\ \hline
\end{tabular}
\caption{Confidence levels of identified lines with automatic assignments during the line search (Fit) and manual assignments during the inspection (MCMC) steps. Contamination types `$\rm C$' are defined in Table \ref{tab:cont}.}
\label{tab:lv}
\end{table*}

\subsection{Inspection and confidence level of the identified \ion{C}{4} absorbers} \label{sec:insp}

A blind detection of \ion{C}{4} is carried out by matching the expected wavelength and absorption strength using an automatic fitting procedure (see details in \citetalias{Li24}).
Then, an initial confidence level of identification is assigned to the absorption lines detected in the automatic line search step, based on their line significance and contamination $\rm C_{p}$.
The definition and usage of $\rm C_{p}$ (and other contamination `$\rm C$' types) are summarized in Table \ref{tab:cont} and Appendix \ref{sec:cont}). 
We assign all lines with $N_{\rm CIV}/\Delta N_{\rm CIV}<3$ a confidence level of $\rm Lv_{\rm fit} = -1$, where $\Delta N_{\rm CIV}$ is the 1$\sigma$ uncertainty of the best fit $N_{\rm CIV}$ of the Voigt profile. 
These lines are not firmly detected and considered as non-detections. 
For detected \ion{C}{4} components, we assign three levels of Lv$_{\rm fit}$ based on their contamination.
We assign a $\rm Lv_{\rm fit} = 3$ to absorption components without contamination, i.e., no $\rm C_{p}$ within $|v| < v_{\text{FWHM}}$, where $v$ is the velocity relative to profile center and $v_{\text{FWHM}}$ is the full width at half maximum of the absorption profile.
We use $\rm Lv_{\rm fit} = 1$ and 2 for the detected high-significance absorption lines with high and low-level contamination, respectively.
We set the contamination levels according to the position and fraction of contaminating pixel numbers, because they typically are the main factors affecting the identification of \ion{C}{4} absorption lines.
We assign a confidence level of $\rm Lv_{\rm fit} = 2$ (i.e., low-level contamination) to significant \ion{C}{4} absorption systems, if there are contaminated pixels $\rm C_{p}$ within the $|v| < v_{\text{FWHM}}$ of modeled \ion{C}{4} line profile.
The confidence level is further reduced to $\rm Lv_{\rm fit} = 1$ (i.e., high-level contamination),
if more than half of the pixels within the $|v| < 0.5v_{\text{FWHM}}$ range are contaminated.

We further inspect and manually adjust the line confidence level of searched \ion{C}{4} absorption components, because the estimation of line significance and contamination level is not always robust in the automatic search step.
The automatic line search introduced in \citetalias{Li24} will likely underestimate the uncertainty of the fitted parameters. We, therefore, apply a Markov Chain Monte Carlo approach (MCMC, using the {\tt emcee} package; \citealt{Foreman13}) to better measure the absorption parameters, including the column density ($N$), the Doppler $b$ factor, and the redshift of the absorption line centroid ($z_{\rm abs}$), together with their associated uncertainties. 
We visually inspect all high-significance absorption lines with $\rm Lv_{\rm fit} \geq 1$, and perform MCMC of Bayesian fitting for components without high-level contamination or where the high-level contamination may be explained by telluric absorption, relic sky emission, adjacent \ion{C}{4} absorption lines within $\pm500 \,\rm km~s^{-1}$ (the velocity separation of the \ion{C}{4} doublet), or absorption contamination from other ions at other redshifts (see Appendix \ref{sec:cont} and Table \ref{tab:cont} for details).

\begin{figure*}
\begin{center}
\includegraphics[width=0.99\textwidth]{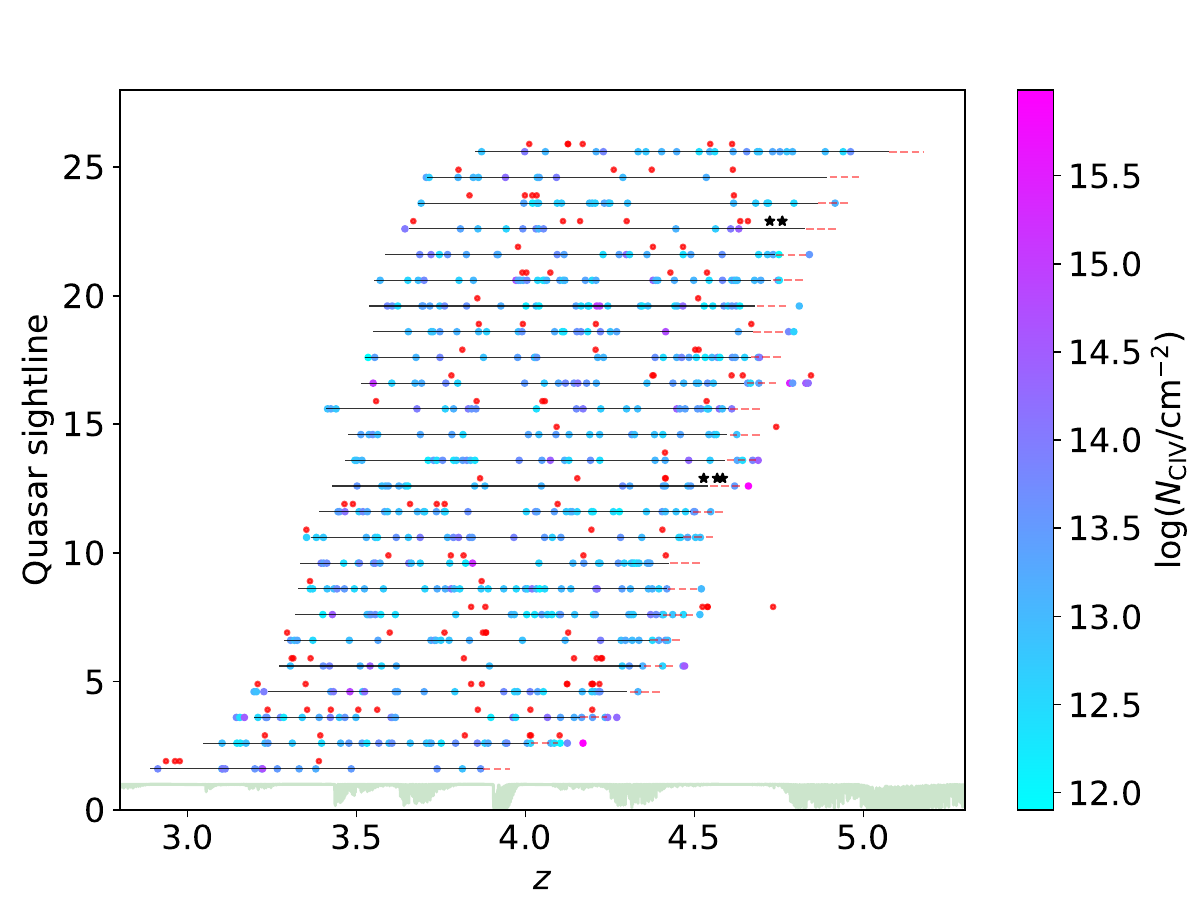}
\caption{Identified \ion{C}{4} absorption lines along each QSO sightline. The black horizontal lines show the intervening regions (defined in Appendix \ref{sec:spec}), while the red dashed lines at the right show the quasar outflow region within 5000 km s$^{-1}$ from $z_{\rm QSO}$, which are excluded in the intervening \ion{C}{4} analyses. Colored points aligned with the horizontal lines show 626 absorption systems constructed from the identified 1263 absorption components at a high confidence level (LV23). In total, there are 557 intervening absorption systems within the intervening redshift regions, excluding BALs represented as black stars above the horizontal lines. 
Red points above horizontal lines are 122 identified absorption components with low confidence levels (LV01), which are not used for constructing absorption systems. The green curve at the bottom shows the telluric absorption contamination.}\label{fig:sys_civ}
\end{center}
\end{figure*}

In the MCMC fitting of the Voigt profile, we typically follow the number of \ion{C}{4} absorption components obtained in the line search step. 
We exclude absorption components without well-fit Voigt models based on visual inspection. These lines are generally weak absorption lines with approximately $3\sigma$ significance measured in the automatic \ion{C}{4} search.
We generally mask the contamination regions if they are not caused by telluric absorption features (e.g., Fig. \ref{fig:lv}). 
In addition to \ion{C}{4} and telluric, we usually include a constant continuum as a free parameter in the model to account for the uncertainties in the continuum, which will be propagated to the measurements of the absorption parameters.

With the MCMC measurements, we assign another set of confidence levels (i.e., $\rm Lv_{MCMC}$) to the inspected absorption lines.
We summarize the assignment of line confidence level $\rm Lv_{fit}$ and $\rm Lv_{MCMC}$ in Table \ref{tab:lv}.
The value of $\rm Lv_{MCMC}$ generally follows $\rm Lv_{fit}$, which assigns the confidence level according to the significance of the line and the contamination, except for two major modifications.
First, we split the high-level contaminated absorption lines with $\rm Lv_{fit} = 1$ into two subclasses, depending on whether the contamination is known or not.
In particular, the confidence level of absorption lines with known contamination is increased to $\rm Lv_{MCMC} = 1.5$, while others are assigned to $\rm Lv_{MCMC} = 1$.
Second, we assign $\rm Lv_{MCMC} = -1$ to all absorption components if they are false-positive detections.  
We provide a detailed description of $\rm Lv_{MCMC}$ assignment and false-positive detection in Appendix \ref{sec:conf}, and show several examples in Fig. \ref{fig:lv}. 
 
We also use the flag `Miss' to mark the missed absorption lines in the automatic search step, which could occur when the fitting is trapped and stops before identifying the desired absorption features.
The missing absorptions are typically associated with complex features in the line search boxes, such as fluctuating relic sky lines and significant blending features.
Similar to the automatically detected features, we apply MCMC measurements to the missed absorption lines and obtain the $\rm Lv_{MCMC}$ value.
Overall, the fraction of missed lines is about $\approx 10\%$ in the full sample.

\subsection{\ion{C}{4} absorption systems}
\label{sec:group}

After the inspection and assignment of confidence level, 
we consider a \ion{C}{4} absorption sample with high confidence levels of $\rm Lv_{MCMC} = 1.5$, 2 and 3  (hereafter the LV23 sample), whereas absorption components with $\rm Lv_{MCMC} = 0$ and 1 are included in the low confidence sample (i.e., the LV01 sample). Then, we adopt a two-step method to construct the sample of \ion{C}{4} absorption systems using the LV23 \ion{C}{4} absorption components.
First, we find all connected absorption features with adjacent absorption components blended with each other, where all fluxes are lower than the continuum by $0.5$ uncertainties.
For each connected absorption feature, we calculate the column-weighted centroids.
Then, we follow the method in \citealt{Davies23a} to iteratively group absorption features with centroid differences smaller than $200~\rm km~s^{-1}$ into a single system.
We repeat this process until no absorption features are separated by less than $200~\rm km~s^{-1}$.

For each absorption system, the redshift is set to the redshift of the component with the highest column density.
We calculate the total column density of absorption systems by adding the column density of grouped absorption components. To avoid the effects of relic sky spikes and blending of absorption systems at other redshifts, the total equivalent width $W_{\rm r}$ is calculated using the Voigt profile model profile as
\begin{align}
W_{\rm r} &= \frac{1}{1+z_{\rm abs}}\sum (1-M_{\rm CIV}(i))\delta\lambda(i),\label{eq:wr}
\end{align}
where $z_{\rm abs}$ is the redshift of absorption line. $M_{\rm CIV}(i)$ and $\delta\lambda(i)$ are the modeled \ion{C}{4} absorption and wavelength interval at pixel $i$, respectively.
The summation is calculated over the $|v| < v_{\rm FWHM}$ region of the modeled \ion{C}{4} doublet. 
We report the lower and upper limits for the total absorption strength using the 16\% and 84\% percentiles of the entire MCMC chains.

We estimate two velocity widths for a given absorption system.
We first estimate the velocity interval enclosing 90 per cent (i.e., from 0.05 to 0.95 cumulative distribution function) of the apparent optical depth profile \citep[e.g.,][]{Savage91, Prochaska97, Davies23a}. 
However, for absorption systems with saturated lines, $\Delta v_{90}$ only reflects the width of the strongest absorption component. Therefore, another estimation of the velocity width $\Delta v_{\rm lc}$ is the velocity difference between absorption lines with the lowest and highest redshifts.

\begin{figure*}
\begin{center}
\includegraphics[width=0.45\textwidth]{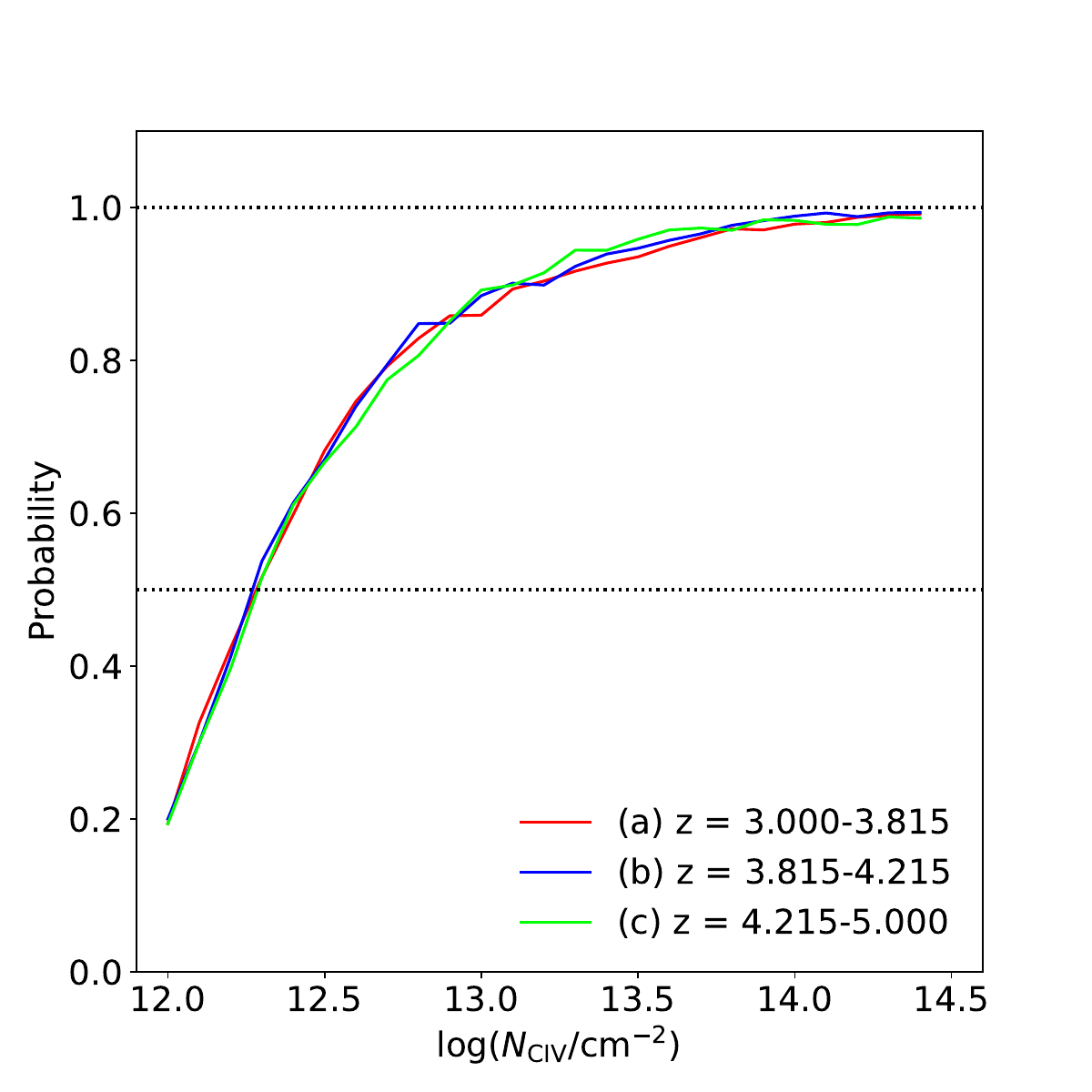}
\includegraphics[width=0.49\textwidth]{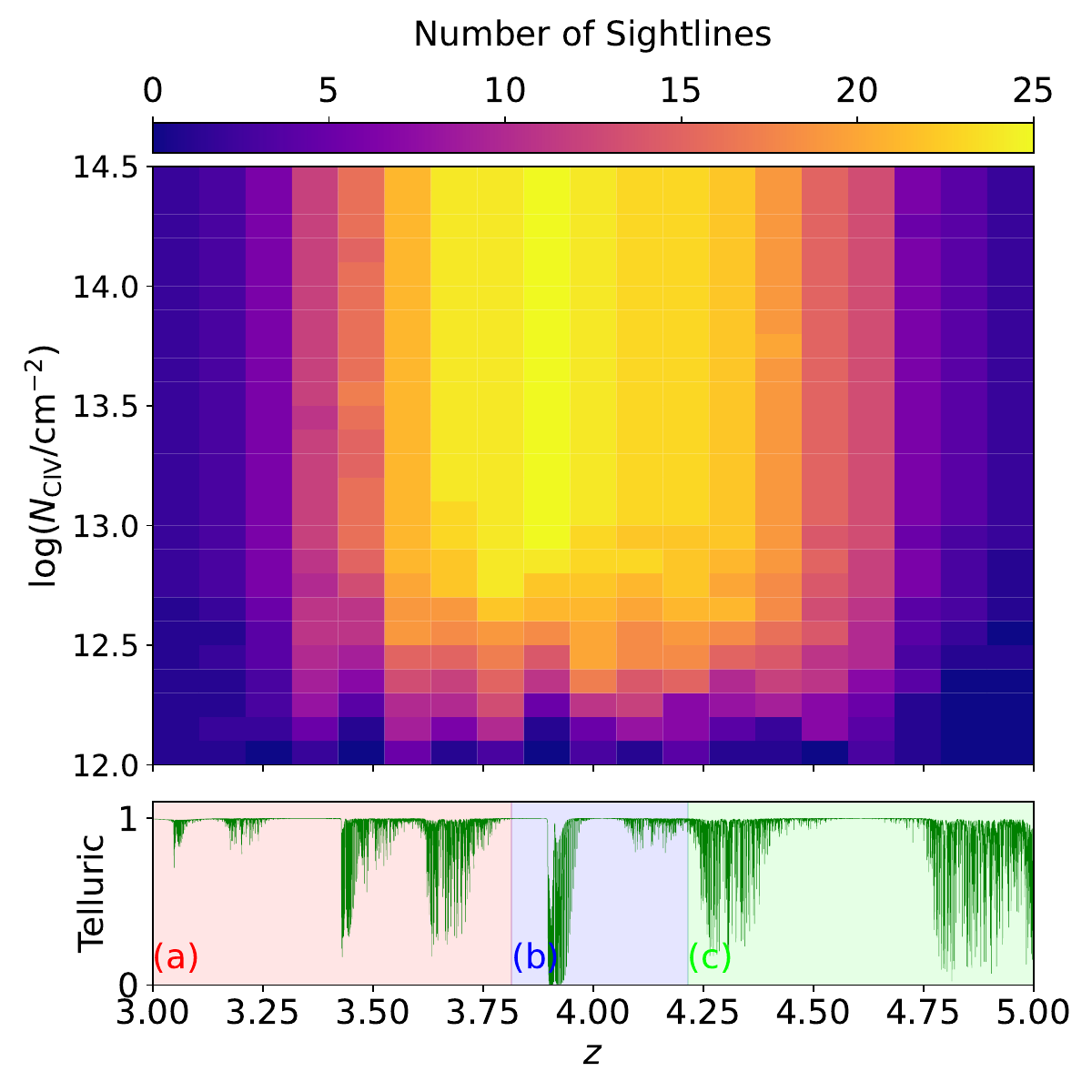}
\caption{Left panel: the completeness curves at different redshift regions for the HIERACHY/MIKE sample. In this work, the completeness curves are calculated using Monte Carlo simulations described in \S\ref{sec:complt}.
Upper right panel: the number of QSO sightlines in the HIERACHY/MIKE sample as a function of redshift and limiting column density.
We define the detection limit as the column density, where the completeness reaches 50\% in the  Monte Carlo simulations.
Lower right panel: the telluric transmission adopted from \citet{Bertaux14} to examine its effects on the detection limit.} 
\label{fig:complt}
\end{center}
\end{figure*}

\subsection{Sample completeness and effective search path}
\label{sec:complt}

The statistical study of \ion{C}{4} absorptions requires knowledge of the completeness $C$ and effective search path $\Delta X_{\text{eff}}$ of the observed \ion{C}{4} absorption line samples.
Sample completeness is influenced by various factors, such as line profiles (e.g., column density and Doppler $b$ factors), data qualities and searching method.
We extract the completeness curve via injecting simulated \ion{C}{4} absorption signals into the intervening \ion{C}{4} search regions of observed quasar spectra (the black horizontal lines in Fig. \ref{fig:sys_civ} and red curves in Fig. \ref{fig:spectra}; see Appendix \ref{sec:spec} for details). 
By using the real spectra, we will obtain the completeness curve by considering the realistic contamination and telluric lines.

\begin{figure*}
\begin{center}
\includegraphics[width=1\textwidth]{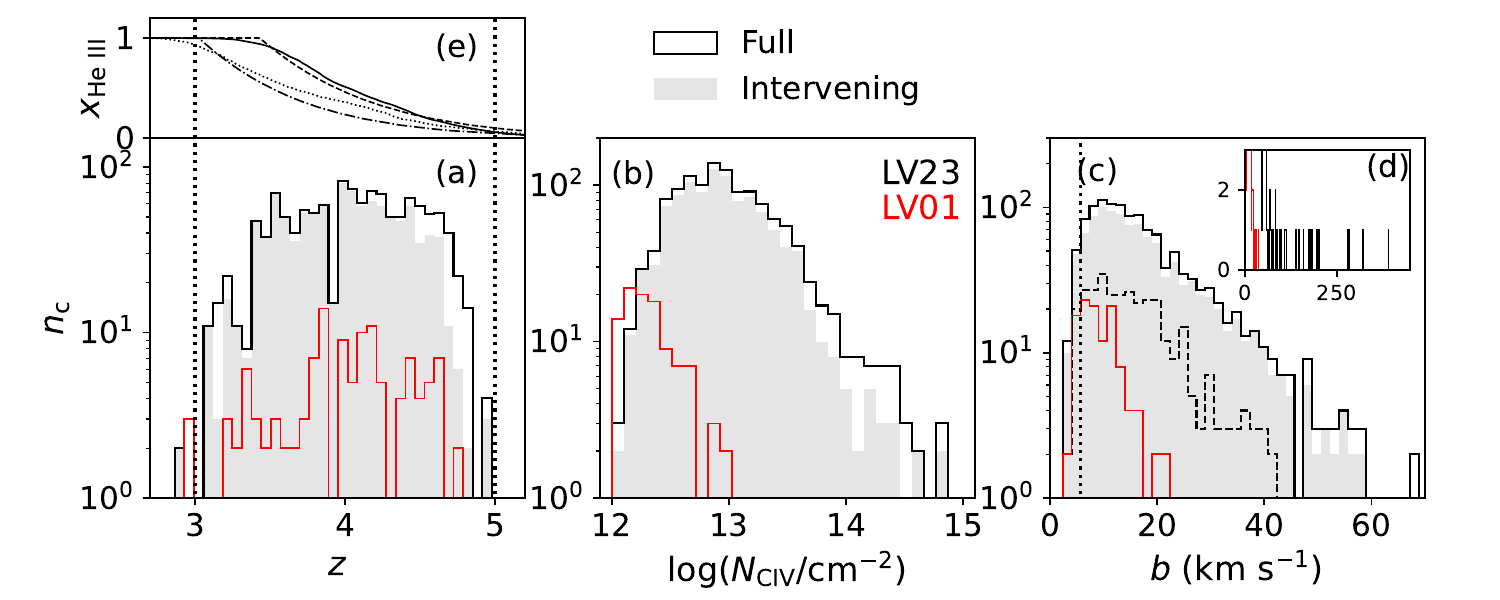}
\caption{The properties of identified \ion{C}{4} absorption components with high (LV23, black) and low (LV01, red) confidence levels (i.e., $\rm Lv_{ MCMC}$). The filled histograms represent intervening LV23 components. Panel (a) shows the redshift distribution. Panel (b) shows the column density distribution. Panel (c) shows the Doppler $b$ parameter distribution with $b = 0-70~\rm km~s^{-1}$, and the inserted Panel (d) shows the full Doppler $b$ range observed in the HIERACHY/MIKE sample.
The black dashed line represents the absorption components in systems with isolated components.
Panel (e) presents the theoretical predictions of \ion{He}{3} ion fraction in the IGM, the dotted line for \citet{McQuinn09}, the solid line for \citet{Compostella14}, the dashed and dotted-dashed lines for \citet{Kulkarni19}. The vertical dotted lines in Panels a and e show the predicted redshift range of \ion{He}{2} EOR, while the vertical dotted line in Panel c indicates the spectral resolution of MIKE quasar spectra.} 
\label{fig:comp_civ}
\end{center}
\end{figure*}

Because of the significant telluric lines (the green curve in Fig. \ref{fig:sys_civ}), the completeness curve depends on the wavelength and varies over the redshift. 
Therefore, we equally split the intervening region of each quasar spectrum into 500 redshift bins.
We then randomly select 50 redshift positions within each redshift bin for the simulated lines.
For a single mock \ion{C}{4} component, we simulated absorption signals using doublet Voigt profiles, which are convolved with the instrument resolution.
The \ion{C}{4} column density ranges from log$N/{\rm cm^{-2}} = 12$ to $14.4$ (about the maximum column in the sample) with a grid of 0.1 dex.
The Doppler $b$ factors of simulated lines are randomly chosen from 8.5, 15.2 and 28.4 $\rm km~s^{-1}$, i.e., close to the median and 1 $\sigma$ percentile of the $b$ distribution for isolated \ion{C}{4} components without adjacent components in the same absorption system (i.e., black dashed histogram in Fig. \ref{fig:comp_civ}). 
We do not use the observed $b$ distribution at specific log$N_{\rm C IV}$ bins for the inserted absorption lines,
because it could be biased since weak absorption lines with broader profiles are more difficult to detect, especially for absorption lines at the detection limit.

Then we perform the \ion{C}{4} doublet search in the simulated spectra and determine the confidence and contamination levels ($\rm Lv_{\rm fit}$), which are the same as the real sample.
The completeness curve is calculated at different redshifts by comparing the detected and the input \ion{C}{4} doublets.
In particular, a successful recovery of the simulated \ion{C}{4} doublet should (1) exhibit a small velocity deviation relative to the inserted lines within $0.5v_{\text{FWHM}}$, and (2) have a $3\sigma$ significance, i.e., $W_{\rm r}/\sigma_{W_{\rm r}} > 3$.
The uncertainties in equivalent width are calculated with 
\begin{align}
\sigma^{2}_{W_{\rm r}} &= \frac{1}{(1+z_{\rm abs})^{2}}\sum E^{2}(i)\delta\lambda^{2}(i)\label{eq:wre},
\end{align}
where $E(i)$ is 1 $\sigma$ flux uncertainty at pixel $i$.
Here, equivalent width is adopted rather than column density because $\sigma_{N}$ is less defined due to the degeneracy between $N$ and $b$.
Then, we obtain detection probability $P(\text{log}N| z)$ for individual QSO sightlines using the simulated line with $\rm Lv_{\rm fit} \geq 1$.

Next, we calculate the sample completeness $C(\text{log}N|z)$ and effective search path $\Delta X_{\rm eff}(\text{log}N|z)$ by summing $P(\text{log}N|z)$ for all QSO sightlines.
For a given redshift bin from $z_{1}$ to $z_{2}$, the comoving path length for each sightline is $\Delta X = X(z_{2}) - X(z_{1})$, where
\begin{align}
X(z) = \frac{2}{3\Omega_{\rm m}}(\sqrt{\Omega_{\rm m}(1+z)^3+\Omega_{\Lambda}}-1),
\end{align}
is valid for cosmologies with $\Omega_{\rm m}+\Omega_{\Lambda} = 1$ \citep{Tytler87}. 
The effective search path length and corresponding completeness of our absorption line sample can then be calculated as
\begin{align}
\Delta X_{\rm eff}(z,\text{log}N) &= \sum_{i=1}^{25} P_{i}(\text{log}N|z) \Delta X, \\
\Delta X_{\rm tot}(z) &= \sum_{i=1}^{25} \Delta X, \\
C(z,\text{log}N) &= \frac{\Delta X_{\rm eff}(\text{log}N|z)}{\Delta X_{\rm tot}(z)}.
\end{align}

The above measured completeness employs simulated lines using a similar identification procedure to that of the observed \ion{C}{4} sample (\S\ref{sec:insp} and \ref{sec:group}), except that we omit steps of system construction and inspection for unknown-contaminated ($\rm Lv_{MCMC} = 1$) and false-positive ($\rm Lv_{MCMC} = -1$) absorption components. This selective omission balances computational efficiency with measurement accuracy, as including these steps would significantly increase simulation and inspection time while introducing only minimal bias. The expected biases from these omissions are smaller than the relative uncertainties in our statistical measurements, which are typically at $\gtrsim0.1$ level as shown in Tables \ref{tab:cddf}, \ref{tab:dnomegadx}, \ref{tab:cddf_comp}, and \ref{tab:dnomegadx_comp}. Specifically, including the unknown-contaminated and false-positive components in the simulated lines contributes a bias of $\lesssim0.1$, as they comprise less than approximately $1\%$ and $10\%$ of identified components, respectively. Additionally, replacing system completeness with the measured (component) completeness for the observed absorption system sample introduces a bias of $\lesssim0.1$ (see Appendix~\ref{sec:bias_comp} for details). Therefore, throughout this paper, we use the measured completeness as the sample completeness for both HIERACHY/MIKE \ion{C}{4} absorption system and component samples.

In the left Panel of Fig. \ref{fig:complt}, we show the sample completeness for the HIERACHY/MIKE sample in several redshift bins.
These redshift bins are the same as that used in the comoving path length number density of \ion{C}{4} $dn_{\rm CIV}/dX$ and the cosmic abundance $\Omega_{\rm CIV}$.
The HIERACHY/MIKE sample achieves completeness 50\% at about log($N_{\rm CIV,det}/\rm{cm}^{-2}) \approx 12.3$, which is deeper than the \citetalias{Hasan20} sample ($W_{\rm r1548,det} \approx 0.05$~\AA) constructed using Keck/HIRES and VLT/UVES quasar spectra and the \citet{Davies23a} sample (log($N_{\rm C IV,det}/\rm cm^{-2})\approx 13.2$) using Xshooter quasar spectra.

In the right Panel of Fig. \ref{fig:complt}, we show the numbers of QSO sightlines with the limiting column density lower than specific column densities as a function of the redshift, where the limiting column density is defined at $P(\text{log}N|z) = 0.5$. At $z\approx3.5-4.5$, the HIERACHY/MIKE sample includes approximately 20 quasars with their column density detection limit at log$(N_{\rm C IV}/\rm cm^{-2})\approx12.3$.

\section{Sample properties and comparison} \label{sec:sample}

We identify 1385 \ion{C}{4} absorption lines using the Magellan/MIKE spectra, including 1263 LV23 (i.e., $\rm Lv_{ MCMC} = 1.5-3$) and 122 LV01 absorption components (i.e., $\rm Lv_{ MCMC} = 0-1$), respectively.
We use these LV23 absorption components to construct 626 \ion{C}{4} absorption systems with the method stated in \S\ref{sec:group}.
The distribution of \ion{C}{4} absorption systems for each QSO sightline is shown as colored dots aligned with horizontal lines in Fig. \ref{fig:sys_civ}.
The colored dots on the black solid horizontal lines represent 557 intervening absorption systems used in \S\ref{sec:stat}, excluding broad absorption lines (BALs, black stars above horizontal lines) as well. 
Some absorption systems exhibit redshifts higher than $z_{\rm QSO}$ (right end of red dashed lines). 
These absorption systems may be due to inflowing gas accreted onto the central supermassive black hole \citep{Zhou19} or their host galaxies \citep{Storchi19}. We also show the LV01 components with red dots above the horizontal lines.

We present the distributions of the redshift, column density, and Doppler $b$ parameter of absorption components for the entire HIERACHY/MIKE sample in Fig. \ref{fig:comp_civ}.
The LV23 sample covers a redshift range of $z\approx3-5$, covering nearly the entire \ion{He}{2} reionization period, and is sensitive to the period where He II reionization is expected to be most active (Panel e, \citealt{McQuinn09, Compostella14, Kulkarni19}). 
The column density peaks at log$(N_{\rm C IV}/\rm cm^{-2})\approx13$, with a minimum value around log$(N_{\rm C IV}/\rm cm^{-2})\approx12$.
The Doppler $b$ parameter has a median value of $b =15.2~\rm kms^{-1}$ with a prominent tail towards $b \sim500~\rm kms^{-1}$.
These broad lines are absorption components of BALs in the spectra of J221111-330245 and J045427-050049.

\begin{figure*}
\begin{center}
\includegraphics[width=1\textwidth]{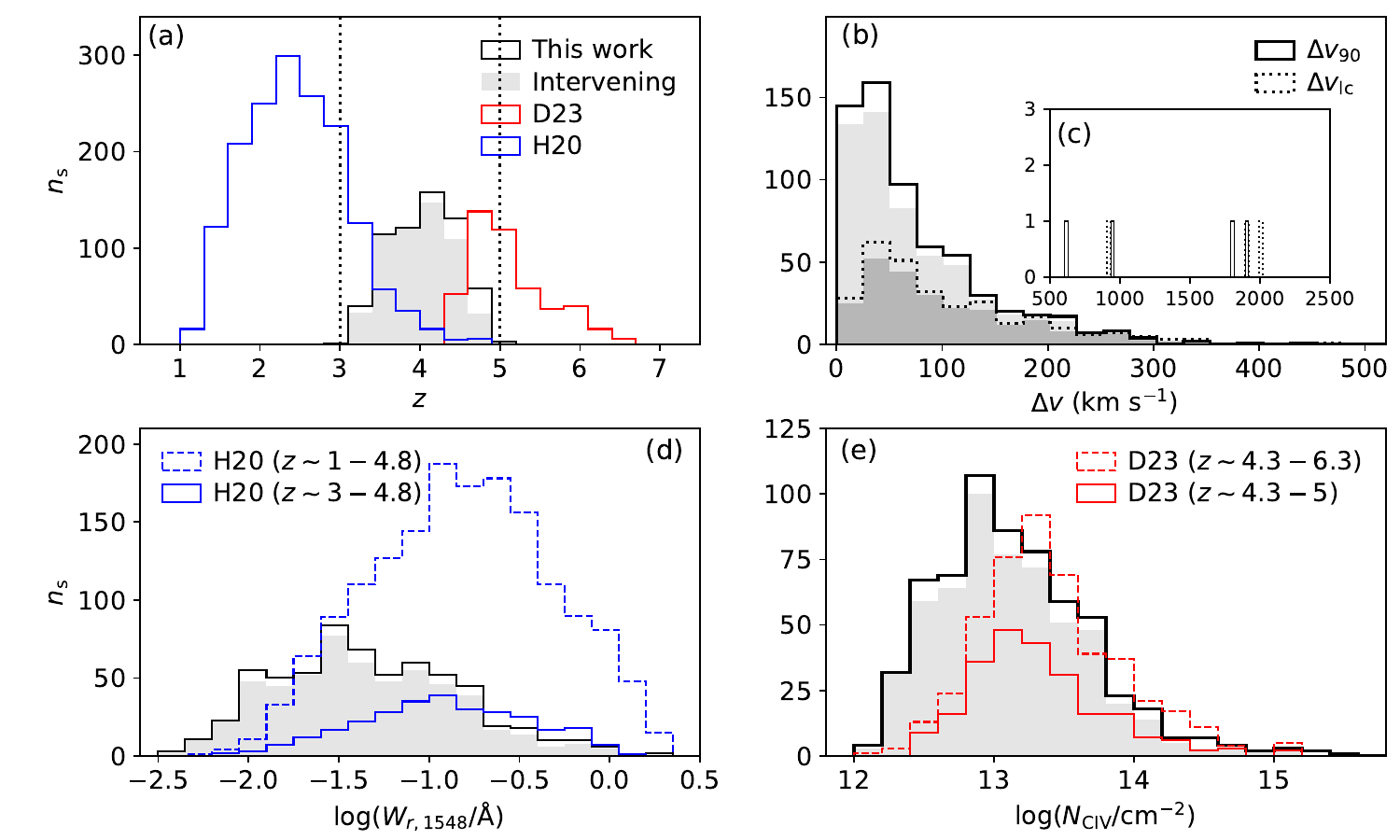}
\caption{The properties of constructed HIERACHY/MIKE \ion{C}{4} absorption systems.
Panel (a): the redshift distribution.
Panel (b-c): the velocity width measure with velocity intervals covering central 90\% optical depth (solid) and between components with lowest and highest redshift (dotted) of absorption systems.
For single-component systems, $\Delta v_{\rm lc}$ is fixed at 0 km s$^{-1}$ and is not plotted in this panel. 
Panel (d): the strength distribution measured with equivalent width.
Panel (e): the column density.
In these panels, The filled histograms represent intervening absorption systems.
We also compare the properties of the HIERACHY/MIKE sample with that of other medium-resolution (red for \citetalias{Davies23b}) and high-resolution (blue for \citetalias{Hasan20}) samples.
The dashed and solid lines for these literature samples represent their full samples and partial samples at $z=3-5$.}
\label{fig:scomp}
\end{center}
\end{figure*}

As shown in Fig. \ref{fig:scomp}, the distribution of absorption systems peaks at $z\approx4$, log$(N_{\rm C IV}/\rm cm^{-2})\approx13$, and $\Delta v_{90}\approx40~\rm km~s^{-1}$, which generally follows the absorption component distribution. These systems also show a long tail in $\Delta v_{90}$ (and $\Delta v_{\rm lc}$) extending to $\approx2000~\rm km~s^{-1}$.
About 340 absorption systems only exhibit single absorption components.
These systems have $\Delta v_{\rm lc} = 0$ and are masked in the Panel b of Fig. \ref{fig:scomp} for clarity. 
Absorption systems with $\Delta v_{90}\gtrsim 500~\rm km~s^{-1}$ are BALs.

The HIERACHY/MIKE \ion{C}{4} absorption system sample is currently the largest sample of weak absorption systems at \ion{He}{2} reionization.
In Fig.\ref{fig:scomp}, we compare our sample with the high-resolution Xshooter sample (\citetalias{Davies23b}; red lines) and KECK/HIRES and VLT/UVES sample in (\citetalias{Hasan20}; blue lines).
Our sample contains more \ion{C}{4} absorbers with log$(N_{\rm CIV}/\rm cm^{-2}) \lesssim 13$ and $W_{\rm r,1548}\lesssim0.1$\AA~at redshift $z=3-5$ (solid lines). Furthermore, the HIERACHY/MIKE sample is also optimized to cover redshift $z\approx4$, where the reionization of \ion{He}{2} to \ion{He}{3} is expected to be most active. 

\section{\ion{C}{4} absorption line statistics} 
\label{sec:stat}

In this section, we study system statistics rather than component statistics, as the latter may be more sensitive to internal dynamics than the cosmic population of absorbers. As described in \S\ref{sec:complt}, we use the component completeness for the observed HIERACHY/MIKE system sample, with the expected bias being smaller than the relative uncertainties measured in this section (also see Appendix \ref{sec:bias_comp}).
We investigate three system statistics: the column density distribution function (CDDF; $f(N)$), comoving path length number density ($dn_{\text{CIV}}/dX$), and cosmic mass abundance ($\Omega_{\text{CIV}}$) of 557 HIERACHY/MIKE intervening \ion{C}{4} absorption systems.
These systems are shown with colored points on black horizontal lines in Fig. \ref{fig:sys_civ} and filled histograms in Fig. \ref{fig:scomp}. Measurements for 1090 intervening components will be presented in Appendix \ref{sec:stat_comp}.

\subsection{Column density distribution function} 
\label{sec:CDDF}

\begin{figure*}
\begin{center}
\includegraphics[width=0.49\textwidth]{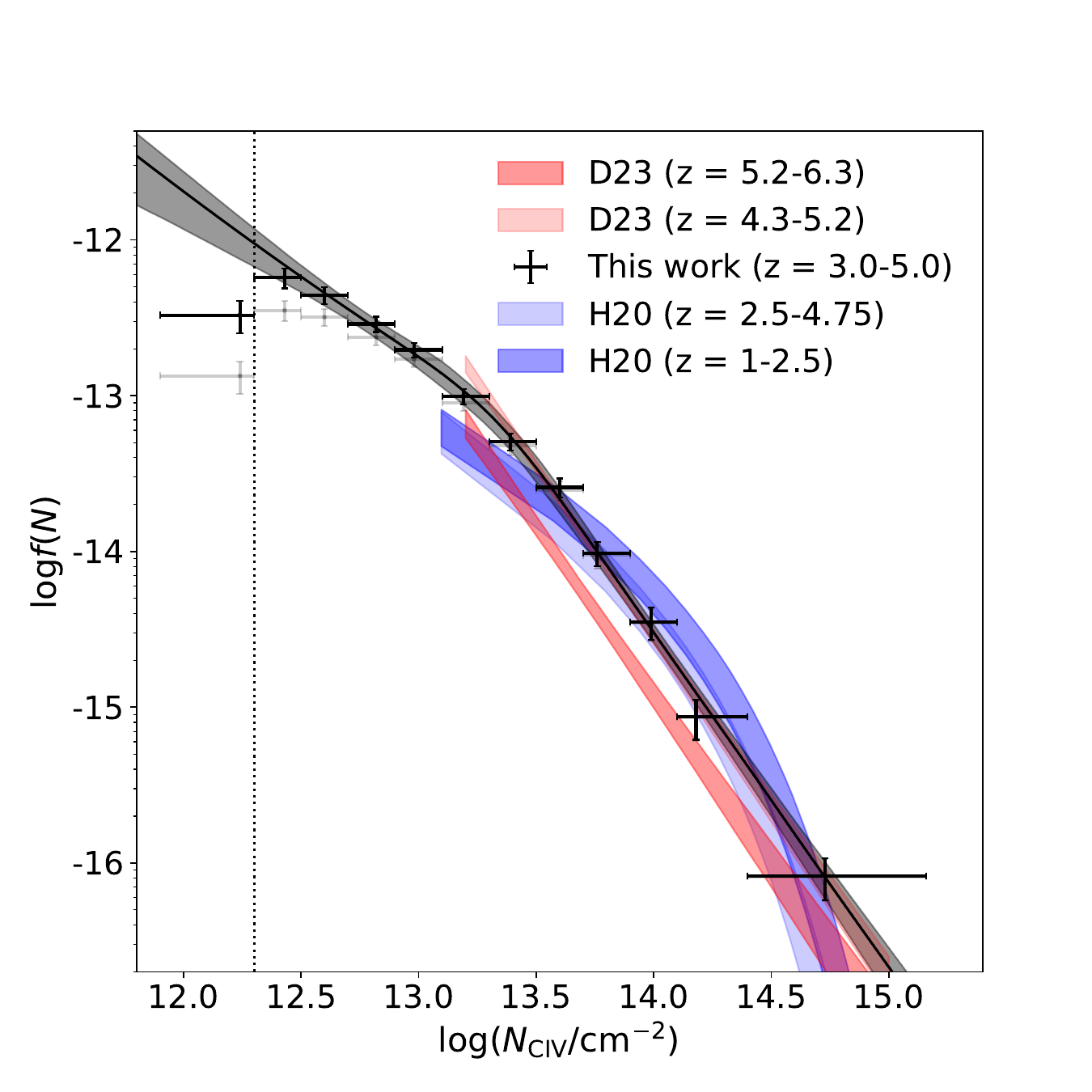}
\includegraphics[width=0.49\textwidth]{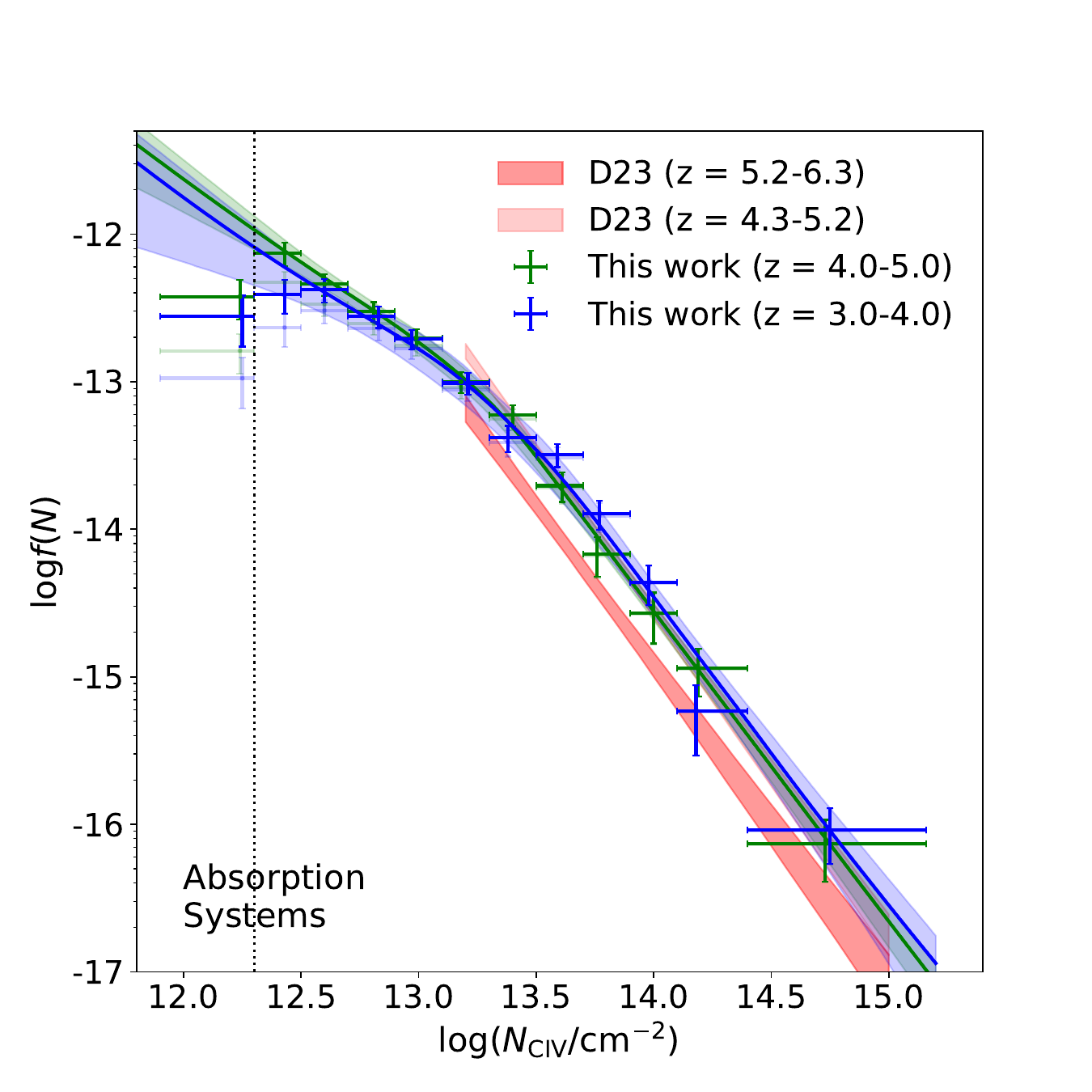}
\caption{The CDDF of the HIERACHY/MIKE \ion{C}{4} absorption systems and their redshift evolution.
Left panel: we compare the CDDF of the full sample (Black dots) with that of \citetalias{Davies23b} and \citetalias{Hasan20}.
Right panel: we show the CDDFs of the HIERACHY/MIKE sample at different redshifts,
compared with \citetalias{Davies23b} sample.
The dark and light data points represent measurements with and without completeness correction, respectively. 
The best-fit solutions of Equation \ref{eq:cddf_fit} for absorption systems with their strength above the 50\% completeness limit (vertical dotted lines) are shown as solid curves, with the corresponding $1\sigma$ uncertainties represented by the shaded regions.} 
\label{fig:cddf}
\end{center}
\end{figure*}

In this work, we adopt the CDDF as the number of absorption systems per unit column density per unit comoving path length $dX$.
Here, we calculate the CDDF of \ion{C}{4} absorption systems in a column density bin log$N_{i}$ with a width of $\Delta N_{i}$ as
\begin{equation}
f(N_{\rm i}) = \sum_{j=1}^{\mathscr{N}_i}\frac{1}{\Delta N_{i}\Delta X_{\rm eff}({\rm log} N_{j}|z_{j})},
\end{equation}\label{eq:cddf}
where $\mathscr{N}_i$ and $\Delta X_{\rm eff}({\rm log} N_{j}|z_{j})$ are the number of absorption systems and the effective search path length for the $j$th absorption system within these bins. 
We split the HIERACHY/MIKE \ion{C}{4} sample into 9 bins at log$(N_{\rm CIV}/\rm cm^{-2})=12.3-14.1$, with a bin width of $\Delta \text{log}N = 0.2$ dex.
We set a single bin for absorption systems with column density log$(N_{\rm CIV}/\rm cm^{-2})<12.3$ ($50\%$ completeness).
The HIERACHY/MIKE sample has a limited number of absorption systems with log$(N_{\rm CIV}/\rm cm^{-2})>14.1$ (Fig. \ref{fig:scomp}).
Therefore, we split these systems into two bins with a boundary at log$(N_{\rm CIV}/\rm cm^{-2})=14.4$.
To investigate the \ion{C}{4} CDDF evolution, we also measure the CDDF of the HIERACHY/MIKE subsample at the redshift ranges of $z\approx 3-4$ and $z\approx 4-5$ (Fig. \ref{fig:cddf}). 
For the CDDF without completeness correction, we simply replace $\Delta X_{\rm eff}({\rm log} N_{i}|z_{j})$ with $\Delta X_{\rm tot}(z_{ j})$.
The $f(N)$ uncertainty is calculated as the Poisson error.

\begin{table*}
\caption{The measured CDDF for the HIERACHY/MIKE \ion{C}{4} absorption systems in different redshift ranges.}
\label{tab:cddf}
\centering
\begin{tabular}{cccccccccccc}
\hline\hline
 & \multicolumn{3}{c}{$ 3.0 < z < 5.0 $} & & \multicolumn{3}{c}{$ 3.0 < z < 4.0 $} & & \multicolumn{3}{c}{$ 4.0 < z < 5.0 $} \\ 
 \cline{2-4} \cline{6-8} \cline{10-12}
 log$N$ range & log$N_{\rm Med}$& $\mathscr{N}$ & log$f(N)$  & &log$N_{\rm Med}$& $\mathscr{N}$ & log$f(N)$  & &log$N_{\rm Med}$&  $\mathscr{N}$ & log$f(N)$ \\ 
\hline
11.9 - 12.3 & 12.24 & 18 & $-12.48 \pm 0.10$ & &12.25 & 7 & $-12.56 \pm 0.16$ & &12.24 & 11 & $-12.42 \pm 0.13$ \\
12.3 - 12.5 & 12.43 & 46 & $-12.24 \pm 0.06$ & & 12.43 & 15 & $-12.41 \pm 0.11$ & & 12.43 & 31 & $-12.13 \pm 0.08$ \\
12.5 - 12.7 & 12.60 & 66 & $-12.36 \pm 0.05$ & & 12.60 & 31 & $-12.38 \pm 0.08$ & & 12.60 & 35 & $-12.34 \pm 0.07$ \\
12.7 - 12.9 & 12.82 & 78 & $-12.54 \pm 0.05$ & & 12.83 & 37 & $-12.56 \pm 0.07$ & & 12.81 & 41 & $-12.52 \pm 0.07$ \\
12.9 - 13.1 & 12.98 & 89 & $-12.71 \pm 0.05$ & & 12.97 & 43 & $-12.71 \pm 0.07$ & & 12.99 & 46 & $-12.70 \pm 0.06$ \\
13.1 - 13.3 & 13.19 & 74 & $-13.00 \pm 0.05$ & & 13.21 & 36 & $-13.01 \pm 0.07$ & & 13.18 & 38 & $-13.00 \pm 0.07$ \\
13.3 - 13.5 & 13.39 & 62 & $-13.29 \pm 0.06$ & & 13.38 & 25 & $-13.38 \pm 0.09$ & & 13.40 & 37 & $-13.23 \pm 0.07$ \\
13.5 - 13.7 & 13.60 & 51 & $-13.59 \pm 0.06$ & & 13.59 & 31 & $-13.49 \pm 0.08$ & & 13.61 & 20 & $-13.70 \pm 0.10$ \\
13.7 - 13.9 & 13.76 & 31 & $-14.01 \pm 0.08$ & & 13.77 & 20 & $-13.89 \pm 0.10$ & & 13.76 & 11 & $-14.17 \pm 0.13$ \\
13.9 - 14.1 & 13.99 & 18 & $-14.45 \pm 0.10$ & & 13.98 & 11 & $-14.36 \pm 0.13$ & & 14.00 & 7 & $-14.57 \pm 0.16$ \\
14.1 - 14.4 & 14.18 & 12 & $-15.06 \pm 0.13$ & & 14.18 & 4 & $-15.23 \pm 0.22$ & & 14.19 & 8 & $-14.94 \pm 0.15$ \\
14.4 - 15.2 & 14.73 & 11 & $-16.08 \pm 0.13$ & & 14.75 & 6 & $-16.04 \pm 0.18$ & & 14.73 & 5 & $-16.13 \pm 0.19$ \\
\hline
\end{tabular}
\end{table*}

The completeness-corrected CDDF of the HIERACHY/MIKE \ion{C}{4} absorption systems shows a broken power-law shape at log$(N_{\rm C IV}/\rm cm^{-2}) \gtrsim 12.3$, where the sample reaches 50\% completeness. 
This CDDF shape is different from that measured at $z\approx 1-4.75$.
In particular, \citetalias{Hasan20} adopted a Schechter function for the KECK/HIRES and VLT/UVES sample, while \citetalias{Davies23b} adopted a single power-law for the Xshooter sample at $z\approx4.3-6.3$, which are both shown in the left Panel of Fig. \ref{fig:cddf}. 
The different column density detection limits are likely to induce this difference, 
as \citetalias{Hasan20} and \citetalias{Davies23b} samples both exhibit 50\% completeness at column density log$(N_{\rm C IV}/\rm cm^{-2})\approx13.2$, about the breakpoint of measured broken power-law CDDF.
The measured CDDF for absorbers with log$(N_{\rm C IV}/\rm cm^{-2})\gtrsim 13.5$ agrees well with that reported at redshift $z\sim 4.3-5.2$ in \citetalias{Davies23b} and $z\sim 2.5-4.75$ in \citetalias{Hasan20}.
We summarize the measured CDDF in Table \ref{tab:cddf}.

To quantitatively investigate the \ion{C}{4} statistical properties, we fit the measured CDDF for absorption systems with their strength above the 50\% completeness limit (i.e., log$(N_{\rm C IV}/\rm cm^{-2}) > 12.3$) by using a broken power-law function as 
\begin{equation}
\text{log}f(N) = 
\begin{cases} 
\text{log}f_{13} + \alpha({\log}N-13) & \text{if } N < N_{\rm crit}, \\ 
\text{log}f_{13} + \beta({\log}N-13)~ + \\
(\alpha-\beta)(\log N_{\rm crit}-13) & \text{if } N \geq N_{\rm crit} ,
\end{cases}
\label{eq:cddf_fit}
\end{equation}
where $\alpha$ and $\beta$ are the power-law slopes below and above $N_{\rm crit}$.
Here, $f_{13}$ is the normalization of $f(N)$ at log$(N_{\rm C IV}/\rm cm^{-2}) = 13$ along the first powerlaw with the slope of $\alpha$, because $\log N_{\rm crit}/{\rm cm^{-2}}$ is higher than 13.
The best-fit solution of $\alpha$, $\beta$, $\text{log}f_{13}$ and $\text{log}N_{\rm crit}$ is obtained under a Bayesian framework with a Markov Chain Monte Carlo (MCMC) approach using the {\tt emcee} package \citep{Foreman13}, which is introduced in details in \citet{Qu2022}.
In this work, the best-fit parameters are obtained using an MCMC chain with 1000 steps and 50 walkers.

For the full sample, the measured slope $\beta$ of CDDF is much steeper than $\alpha$, i.e. $\beta = -2.16^{+0.16}_{-0.14}$ and $\alpha = -1.04^{+0.20}_{-0.26}$, with the critical column density of log$(N_{\rm crit}/{\rm cm^{-2})} = 13.35^{+0.20}_{-0.19}$. 
The fitted results are summarized in Table \ref{tab:tb_cddf_fit}, and the best-fit models are compared with measurements in Fig. \ref{fig:cddf}, where the shaded regions represent the $1\sigma$ confidence level.
The CDDF functions show little evolution between the two redshift bins (i.e., $z\approx 3-4$ and $z\approx 4-5$), with all best-fit parameters consistent within $1\sigma$.

\subsection{Comoving path length number density} 
\label{sec:dndx}

Next, we examine the evolution of the comoving path length number density of \ion{C}{4} absorbers $dn_{\rm CIV}/dX$, which is defined as the number of absorption systems per unit comoving path length ($X$).
We measure $dn_{\rm CIV}/dX$ for MIKE \ion{C}{4} absorption systems by dividing the entire sample into three redshift bins, with the redshift ranges summarized in Table \ref{tab:dnomegadx}. 
Results based on bins with their boundaries set by variable telluric contamination are presented in Appendix \ref{sec:zbin_tel}.
In a redshift bin, we follow \citetalias{Hasan20} and measure $dn_{\rm CIV}/dX$ and the associated uncertainties $\sigma^{2}_{dn/dX}$ within the $i$th redshift bin $z_{i}$ as
\begin{align}
dn/dX &= \sum_{j=1}^{\mathscr{N}_{i}}\frac{1}{\Delta X_{\rm eff}(\text{log}N_{j}|z_{i})}, \\
\sigma^{2}_{dn/dX} & = \sum_{j=1}^{\mathscr{N}_{i}}(\frac{1}{\Delta X_{\rm eff}(\text{log}N_{j}|z_{i})})^{2},
\end{align}
where $\mathscr{N}_{i}$ is the number of absorption systems with $\log N_{\rm CIV}$ in a given column density bin, and $j$th absorption system has a column density of $\text{log}N_{j}$.

\begin{table}
\caption{Best-fit models of CDDF for the HIERACHY/MIKE \ion{C}{4} absorption systems.}\label{tab:tb_cddf_fit}
\centering
\renewcommand{\arraystretch}{1.3}
\begin{tabular}{ccccc}
\hline
\hline
$z$ range & log$N_{\rm crit}$ & log$f_{13}$ & $\alpha$ & $\beta$ \\ 
\hline
3.0-5.0 &$13.35^{+0.20}_{-0.19}$& $-12.77^{+0.06}_{-0.09}$& $-1.04^{+0.20}_{-0.26}$& $-2.16^{+0.16}_{-0.14}$ \\ \hline
3.0-4.0 &$13.42^{+0.34}_{-0.30}$& $-12.80^{+0.11}_{-0.17}$& $-0.96^{+0.33}_{-0.47}$& $-2.15^{+0.34}_{-0.25}$ \\ \hline
4.0-5.0 &$13.31^{+0.22}_{-0.13}$& $-12.75^{+0.06}_{-0.10}$& $-1.09^{+0.18}_{-0.27}$& $-2.13^{+0.12}_{-0.14}$ \\ \hline
\end{tabular}
\end{table}

\begin{figure*}
\begin{center}
\includegraphics[width=0.49\textwidth]{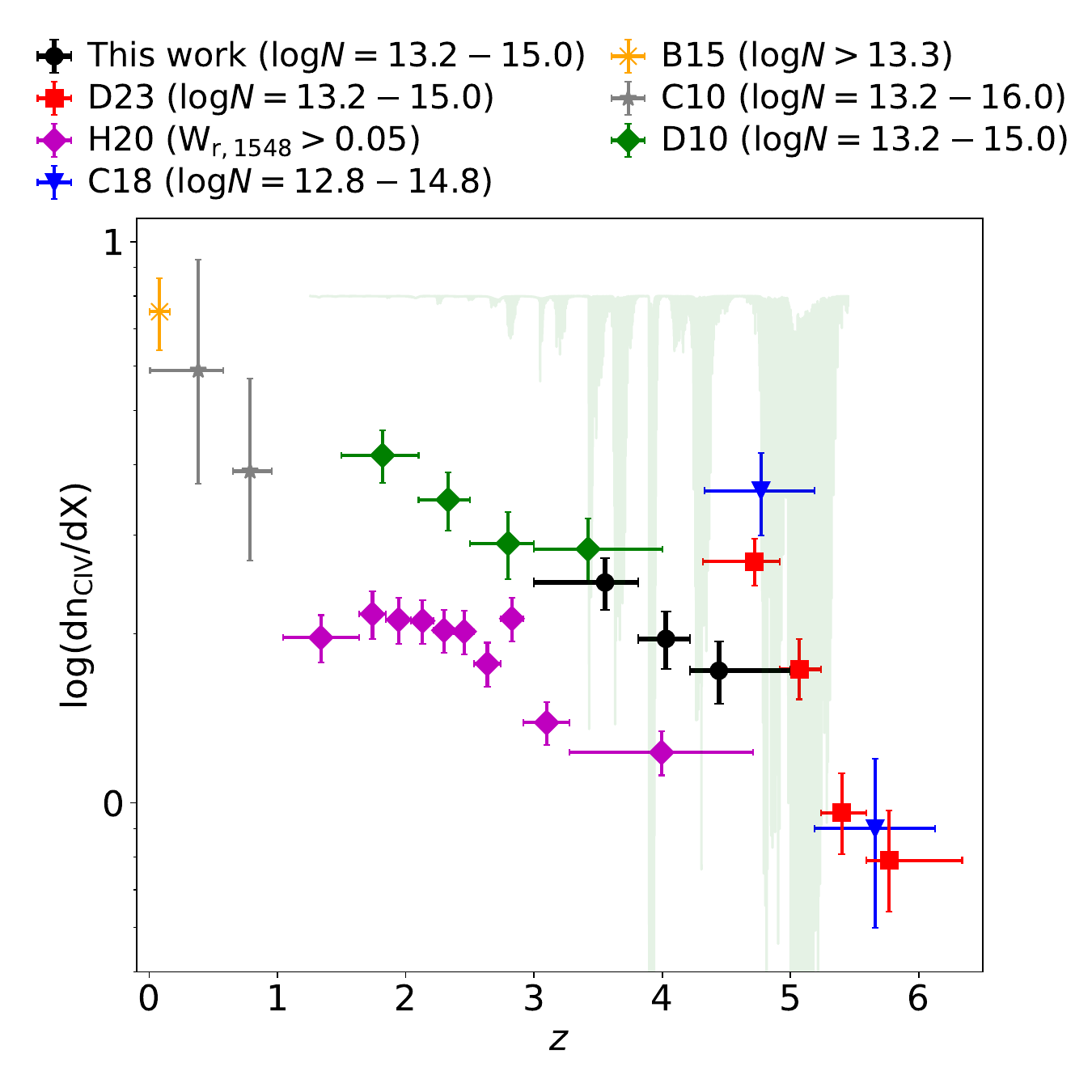}
\includegraphics[width=0.49\textwidth]{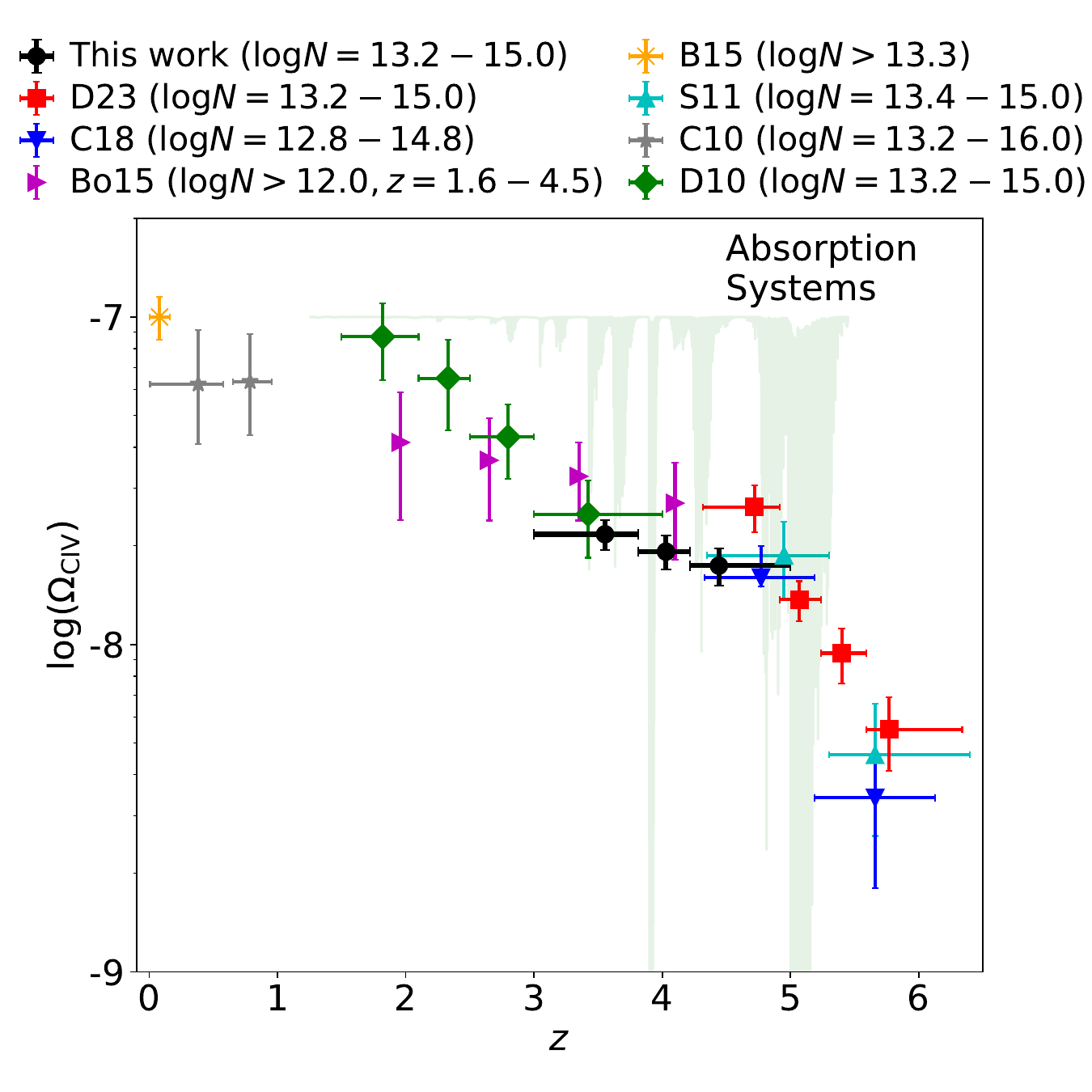}
\caption{The evolution of \ion{C}{4} absorption systems over $z\approx0-6$.
Left panel: the comoving path length number density $dn_{\rm CIV}/dX$ of the HIERACHY/MIKE and literature samples. 
Right panel: the cosmic mass abundance of \ion{C}{4} $\Omega_{\rm CIV}$. 
These literature samples include C10 \citep{Cooksey10}, D10 (\citealt{D'Odorico10}'s data recalculated in \citetalias{Davies23b}), S11 \citep{Simcoe11}, B15 \citep{Burchett15}, Bo15 \citep{Boksenberg15}, C18 \citep{Codoreanu18}, \citetalias{Hasan20}, and \citetalias{Davies23b}. The column density limits used for these measurements are labeled in the legend.
The green shaded regions represent the telluric features.} 
\label{fig:evol_com}
\end{center}
\end{figure*}

We first measure the $dn_{\rm CIV}/dX$ for absorption systems with log$(N_{\rm C IV}/\rm cm^{-2})\gtrsim 13.2$ and compare them with previous results showing similar column density ranges in the left Panel of Fig. \ref{fig:evol_com}.
The measured $dn_{\rm CIV}/dX$ show a $\approx 2.2\sigma$ increase as the redshift decreases over $z\approx 3-5$.
Our measurement $dn_{\rm CIV}/dX_{z\approx 4.4}$ of $1.72\pm 0.22$ is lower than the measurement of $dn_{\rm CIV}/dX_{z\approx4.7} = 2.69^{+0.33}_{-0.39}$ in \citetalias{Davies23b}, where the difference is about $2\sigma$.
However, the $dn_{\rm CIV}/dX$ evolution trend of the HIERACHY/MIKE sample could naturally extend to their measured values at $z$ range without significant telluric contamination (green shaded region).
Therefore, we suggest the difference between our measurements and \citetalias{Davies23b} is due to different treatments of the completeness correction.

The $dn_{\rm CIV}/dX_{z\approx 3.6}$ of $2.47\pm 0.26$ is consistent with \citealt{D'Odorico10} data of $dn_{\rm CIV}/dX_{z\approx 3.4} = 2.83\pm0.38$ recalculated in \citetalias{Davies23b}, but significantly higher than $dn_{\rm CIV}/dX_{z\approx 3.1} = 1.39\pm0.12$ reported in \citetalias{Hasan20}.
We noticed that the reported $dn_{\rm CIV}/dX$ with limiting $W_{r, 1548} = 0.05$\AA~ in \citetalias{Hasan20} are significantly lower than all other samples with limiting $\log N_{\rm CIV}/{\rm cm^{-2}}\approx 13.2$.
Assuming a typical Dopper $b\approx 10-20\rm~km~s^{-1}$ (Fig. \ref{fig:comp_civ}), $W_{r, 1548} = 0.05$\AA~ is comparable to $\log N_{\rm CIV}/{\rm cm^{-2}}\approx 13.2$, therefore the reason of the lower $dn_{\rm CIV}/dX$ in \citetalias{Hasan20} is still uncertain.

We also study the evolution of $dn_{\rm CIV}/dX$ for absorption systems with different strengths, as previous studies indicate a dependence of $dn_{\rm CIV}/dX$ evolution on absorption strength.
In particular, we adopt the column density bins as log$(N_{\rm C IV}/\rm cm^{-2}) = 12.3-13.2$ (weak), $13.2-13.8$  (moderate), and $13.8-15.2$ (strong).
The $dn_{\rm CIV}/dX$ measurements of moderate and strong absorption samples exhibit a significant increase with decreasing redshift.
In contrast, the weak sample is constant over $z\approx 3-5$ (the left Panel of Fig. \ref{fig:evol_nbin}).
Additionally, over the $z\approx 3-5$, $dn_{\rm CIV}/dX$ of weak absorbers is always larger than that of stronger absorbers, which is also suggested in previous studies (e.g., \citetalias{Hasan20}).

\begin{table}
\caption{The $dn_{\rm CIV}/dX$ and $\Omega_{\rm CIV}$ for the HIERACHY/MIKE \ion{C}{4} absorption systems in different column density ranges.}\label{tab:dnomegadx}
\centering
\begin{tabular}{cccccc}
\hline\hline
$z_{\rm Med}$ & $z$ range & $\Delta X$ & $\mathscr{N}$ & $dn/dX$ & $\Omega_{\rm CIV} (10^{-8})$  \\ \hline
\multicolumn{6}{c}{$ 12.3 < \log (N/{\rm cm}^2) < 13.2 $} \\ \hline
3.600 &3.000-3.815 & 25.4 & 110 & $3.74\pm0.36$ & $0.37\pm0.04$ \\ \hline
4.052 &3.815-4.215 & 33.6 & 98 & $3.33\pm0.34$ & $0.33\pm0.03$ \\ \hline
4.427 &4.215-5.000 & 24.3 & 110 & $3.81\pm0.37$ & $0.40\pm0.04$ \\ \hline
\multicolumn{6}{c}{$ 13.2 < \log (N/{\rm cm}^2) < 13.8 $} \\ \hline
3.560 &3.000-3.815 & 34.6 & 68 & $1.95\pm0.24$ & $0.93\pm0.11$ \\ \hline
4.029 &3.815-4.215 & 36.3 & 55 & $1.54\pm0.21$ & $0.74\pm0.10$ \\ \hline
4.457 &4.215-5.000 & 35.4 & 47 & $1.36\pm0.20$ & $0.59\pm0.09$ \\ \hline
\multicolumn{6}{c}{$ 13.8 < \log (N/{\rm cm}^2) < 15.2 $} \\ \hline
3.502 &3.000-3.815 & 36.5 & 20 & $0.55\pm0.12$ & $1.66\pm0.37$ \\ \hline
4.019 &3.815-4.215 & 37.7 & 17 & $0.45\pm0.11$ & $1.75\pm0.42$ \\ \hline
4.415 &4.215-5.000 & 36.0 & 13 & $0.36\pm0.10$ & $1.15\pm0.32$ \\ \hline
\multicolumn{6}{c}{$ 13.2 < \log (N/{\rm cm}^2) < 15.0 $} \\ \hline
3.554 &3.000-3.815 & 34.6 & 87 & $2.47\pm0.26$ & $2.17\pm0.23$ \\ \hline
4.029 &3.815-4.215 & 37.7 & 71 & $1.96\pm0.23$ & $1.92\pm0.23$ \\ \hline
4.442 &4.215-5.000 & 35.4 & 60 & $1.72\pm0.22$ & $1.74\pm0.22$ \\ \hline
\end{tabular}
\end{table}

\subsection{Cosmic mass abundance} 
\label{sec:omega}

In addition to the comoving path length number density, we also calculate the cosmic mass abundance of \ion{C}{4}, defined as the \ion{C}{4} mass per unit comoving Mpc, in the units of critical density $\rho_{\text{crit}}$ of Universe at $z\sim0$.
\begin{align}
\Omega_{\rm CIV} &= \rho_{\rm CIV}/\rho_{\rm crit}, \\
&=\frac{H_0 m_{\rm C}}{c\rho_{\rm crit}} \int N f(N) dN.
\end{align}
In cases where the sample size of absorption systems is too small to recover the CDDF,  \citet{Storrie96} proposed an estimation of $\Omega_{\rm CIV}$ as 
\begin{equation}\label{eqn:pciv_approx}
 \Omega_{\rm{C}\textsc{IV}} \simeq \frac{H_0 m_{\rm C}}{c \rho_{\rm crit}} \frac{\Sigma N}{\Delta X}.
\end{equation}
This approximation is usually used in previous studies with limited sample sizes (e.g., \citealt{Ryan-Weber09,D'Odorico13,Diaz16,Meyer19,D'Odorico2022}; \citetalias{Davies23b}).
We also use this approximation to study the redshift evolution of $\Omega_{\rm CIV}$,
because the HIERACHY/MIKE sample does not have enough absorption systems (especially log$N_{\rm CIV}/\rm{cm}^{-2} \gtrsim 13.5$) to recover the broken power-law CDDF (Table \ref{tab:cddf}). 

The numbers of redshift and column density bins for $\Omega_{\rm CIV}$ are the same as those used in the measurement of $dn_{\rm CIV}/dX$ (Table \ref{tab:dnomegadx}).
We measure $\Omega_{\rm CIV}$ in the redshift bin $z_i$, 
\begin{align}
 \Omega_{\rm{CIV}}(z_i) \simeq \frac{H_0 m_{\rm C}}{c \rho_{\rm crit}} \sum_{j=1}^{\mathscr{N}_{i}}\frac{N_{j}}{\Delta X_{\rm eff}(\text{log}N_{j}|z_{i})}.
\end{align}
The uncertainty of $\Omega_{\rm CIV}$ is calculated using the fractional variance introduced in \citep{Storrie96}:
\begin{align}
 (\frac{\sigma_{\Omega_{\rm{CIV}}}}{\Omega_{\rm{CIV}}})^{2}(z_i) \simeq \frac{\sum_{j=1}^{\mathscr{N}_{i}} N_{j}^{2}}{(\sum_{j=1}^{\mathscr{N}_{i}}N_{j})^{2}}.
\end{align}
All measured cosmic \ion{C}{4} abundances and uncertainties are are summarized in Table \ref{tab:dnomegadx}.

In the right Panel of Fig. \ref{fig:evol_com}, we present the evolution of $\Omega_{\rm CIV}$ for absorption systems, together with literature results.
The evolution of cosmic mass abundance $\Omega_{\rm CIV}$ generally follows the evolution trend of $dn_{\rm CIV}/dX$ described in \S\ref{sec:dndx}. 
The $\Omega_{\rm CIV}$ shows a $\approx1.4\sigma$ increase as the redshift decreases from $z\approx5$ to 3.
This is consistent with the previous suggestion that an initial significant increase in $\Omega_{\rm CIV}$ after the end of \ion{H}{1} reionization, followed by a shallower increase at $z\lesssim5$ (e.g., \citealt{Simcoe11}; \citetalias{Davies23b}).

\begin{figure*}
\begin{center}
\includegraphics[width=0.49\textwidth]{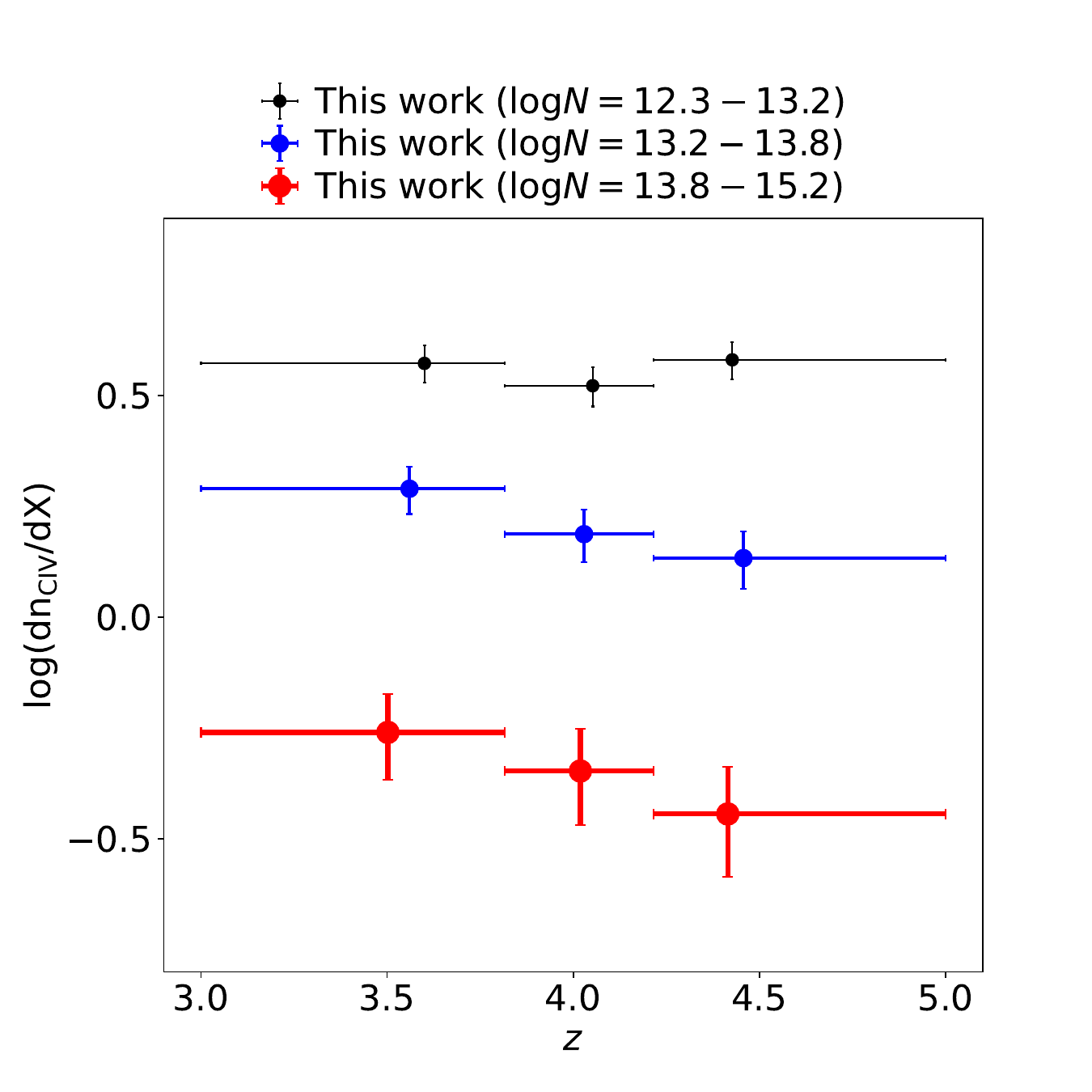}
\includegraphics[width=0.49\textwidth]{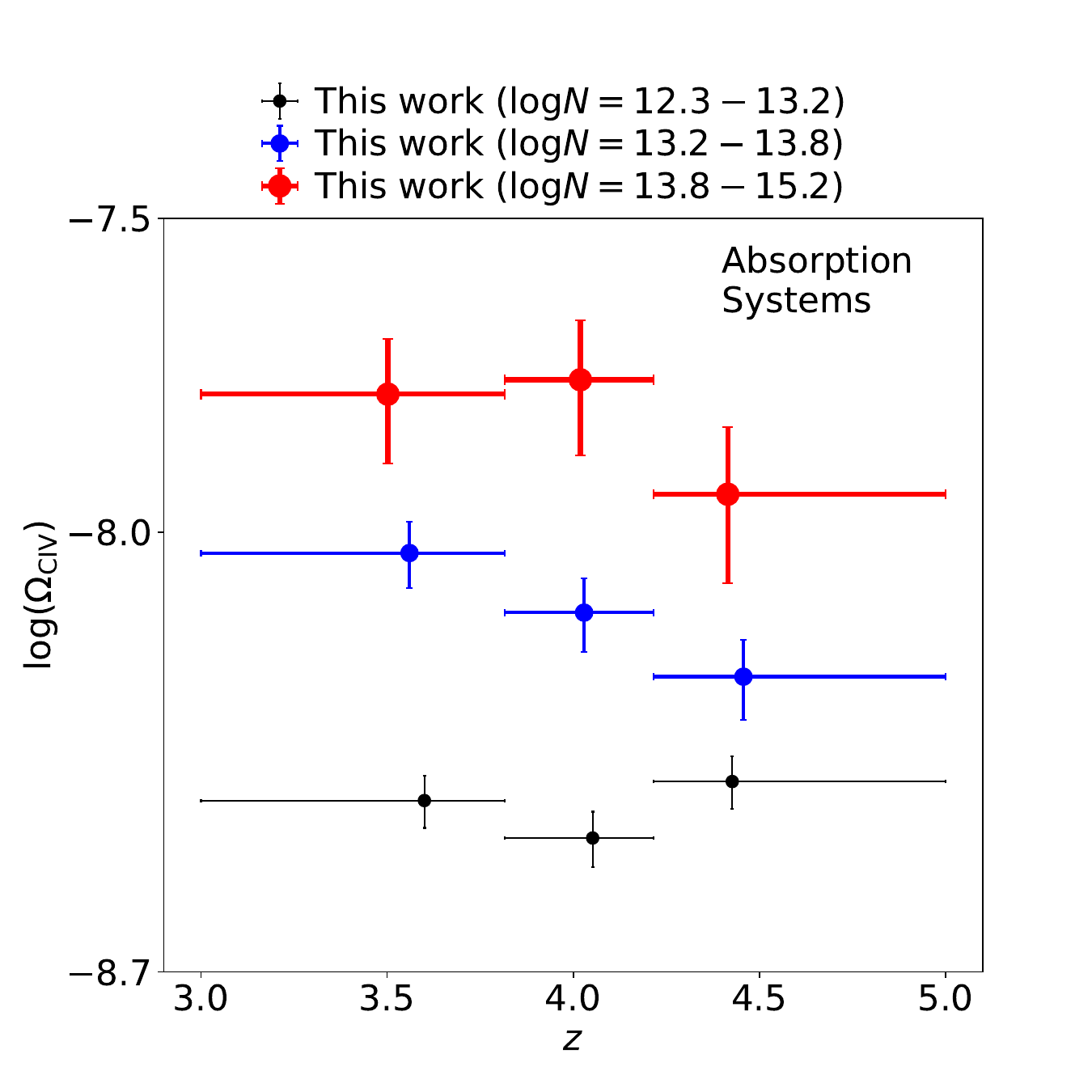}
\caption{The evolution of cosmic \ion{C}{4} statistics for HIERACHY/MIKE absorption systems in different column density ranges: $dn_{\rm CIV}/dX$ (left panel) and $\Omega_{\rm CIV}$ (right panel).
} 
\label{fig:evol_nbin}
\end{center}
\end{figure*}

Similar to $dn_{\rm CIV}/dX$, we also examine the evolution of $\Omega_{\rm CIV}$ in different column density bins in the right Panel of Fig. \ref{fig:evol_nbin}. 
The measurements of $\Omega_{\rm CIV}$ of weak absorption systems do not exhibit significant cosmic evolution, which is also consistent with $dn_{\rm CIV}/dX$.

\section{The HIERACHY/MIKE \ion{C}{4} Catalog} 
\label{sec:catalog}
 
The complete catalog of observed HIERACHY/MIKE \ion{C}{4} absorption sample will be published as supplementary material with this article, with the description of catalog columns shown in Table \ref{tab:obs_desc}. This catalog contains \ion{C}{4} absorption components (systems) with all $\rm Lv_{ MCMC}$ confidence levels.
For convenience, we also report the quasar properties, observation information, and data quality.
In this catalog, we set all unavailable values to be -99. 

First, for each QSO, we report the QSO name, Ra, Dec, Redshift, and the reference of redshift measurement in columns `QSO', `Ra', `Dec',  `z\_QSO', and `z\_ref', respectively. 
In addition, the QSO magnitudes $i$, $J$, and $R_{\rm p}$ are reported in columns `Mag\_i', `Mag\_J', and `Mag\_Rp'.

Then, for the observation information, we report the observation date, total exposure time, slit type, and typical seeing in the columns `Date\_obs', `Exptime', `Slit', and `Seeing', respectively.
Furthermore, the column `zrCIV' shows the intervening redshift ranges defined in Appendix \ref{sec:spec} and used in \S\ref{sec:stat}, while the associated 25 and 75 percentiles of the SN within the intervening \ion{C}{4} search regions are reported in 
columns `S/N25\_zrCIV' and `S/N75\_zrCIV', respectively.

Third, for each detected \ion{C}{4} absorption component, we reported the assigned confidence level and measurements from Voigt profile fitting.
Every \ion{C}{4} component exhibits a unique absorption ID, shown in the column `Id\_abs'. 
The absorption ID is composed of three parts: the QSO field, the system ID, and the component ID.
In particular, the QSO ID is assigned as `q'.
Then, the systems are sorted from low redshift to high redshift, denoted `s' together with an index starting from 1.
Similarly, within each system, individual components are also sorted with increasing redshift, denoted as `c' followed by numbers.
For example, `q1205-0742s1c1' is the first absorption component in the first system in the sightline of J120523-074232.
For each component, we report the confidence level in the column `LV\_MCMC', as introduced in Appendix \ref{sec:conf} and Table \ref{tab:lv}.
Then columns `z\_comp' and `b\_comp' represent the redshift and Doppler $b$ parameter of the \ion{C}{4} absorption component, respectively.
The associated uncertainties are shown in columns `z\_err\_comp' and `b\_err\_comp'.
We report the lower ($16\%$) and upper ($84\%$) limits for the column density of absorption components `logN\_comp' in the `logN\_16\_comp' and `logN\_84\_comp' columns, respectively.
In this paper, we define the uncertainties of `logN\_comp' as the deviation between `logN\_comp' and `logN\_16\_comp'.

Last, for each \ion{C}{4} absorber, we report its redshift, total column density, and corresponding equivalent width in columns `z\_sys', `logN\_sys', `Wr1548\_sys', and `Wr1550\_sys', respectively.
The system redshift is assigned as the component redshift with the highest $\log N_{\rm CIV}$.
The total column density and equivalent width are calculated as the summation of all components with $\rm Lv_{ MCMC} = 1.5, 2, 3$.
We report the lower ($16\%$) and upper ($84\%$) limit for total column density and equivalent widths in columns `logN\_16\_sys', `Wr1548\_16\_sys', and `Wr1550\_16\_sys', `logN\_84\_sys', `Wr1548\_84\_sys', and `Wr1550\_84\_sys', respectively.
In addition to absorption strength, we also report the kinematic properties. In columns `dv\_tau90' and `dv\_lc', we report the velocity widths of absorption systems $\Delta v_{90}$ and $\Delta v_{\rm lc}$, as introduced in \S\ref{sec:group}.
In addition, we also report the local SN of the continuum in the column `S/N\_abs', which is the median S/N for continuum pixels within the $\pm 250\rm~km~s^{-1}$ region around given absorption systems.
The column `Miss' shows whether a absorption component is missed in the automatic line search step (e.g., `Y' or `N').
We report the type of false-positive detection in column `False\_positive'.
Finally, we leave comments in the column `comment' for absorption components, which generally show information such as whether the components are BALs (e.g., `BAL') or are strongly affected by telluric (e.g., `tel') and sky emission (e.g., `sky'), and so on.

\begin{longtable*}{l>{\raggedright\arraybackslash}p{15cm}}
\caption{Descriptive table for the HIERACHY/MIKE \ion{C}{4} absorption line catalog.}\label{tab:obs_desc} \\
\hline
\hline
Column Name & Description \\
\hline
\endfirsthead

\multicolumn{2}{c}%
{{\bfseries \tablename\ \thetable{} -- continued from previous page}} \\
\hline
Column Name & Description \\
\hline
\endhead

\hline
\multicolumn{2}{r}{{Continued on next page}} \\
\endfoot

\hline
\endlastfoot
QSO & Quasar name \\ \hline
Ra & Right Ascension of the quasar (hh:mm:ss) \\ \hline
Dec & Declination of the quasar (dd:mm:ss) \\ \hline
z\_QSO & Redshift of the quasar \\ \hline
z\_ref & Redshift reference \\ \hline
Mag\_i & i-band magnitude of the quasar \\ \hline
Mag\_J & J-band magnitude of the quasar \\ \hline
Mag\_Rp & Rp-band magnitude of the quasar \\ \hline
Date\_obs & Date of the observation \\ \hline
Exptime & Total exposure time (hours) \\ \hline
Slit & Used slit (${\prime\prime}\times{\prime\prime}$) \\ \hline
Seeing & Typical seeing (${\prime\prime}$) \\ \hline
zrCIV & Redshift range of the CIV intervening region \\ \hline
S/N25\_zrCIV & 25th percentile of signal-to-noise ratio within the CIV intervening region \\ \hline
S/N75\_zrCIV & 75th percentile of signal-to-noise ratio within the CIV intervening region \\ \hline
Id\_abs & Identifier for the absorbers, format is q+hhmm+ddmm+s$i$+c$j$ \\ \hline
LV\_MCMC & Line confidence level \\ \hline
z\_comp & Redshift of component \\ \hline
z\_err\_comp & Redshift error of component \\ \hline
logN\_comp & Logarithmic column density of absorption component \\ \hline
logN\_16\_comp & 16th percentile of component logarithmic column density. In this paper, we define the uncertainties of the column density as logN\_comp - logN\_16\_comp \\ \hline
logN\_84\_comp & 84th percentile of component logarithmic column density \\ \hline
b\_comp & Doppler b-factor of component ($\rm km~s^{-1}$) \\ \hline
b\_err\_comp & Doppler b-factor error of component ($\rm km~s^{-1}$) \\ \hline
z\_sys & System redshift \\ \hline
logN\_sys & Total logarithmic column density of CIV absorption system \\ \hline
logN\_16\_sys & 16th percentile of total logarithmic column density of absorption system \\ \hline
logN\_84\_sys & 84th percentile of total logarithmic column density of absorption system \\ \hline
Wr1548\_sys & Total modeled equivalent width of CIV1548 absorption system (\AA) \\ \hline
Wr1548\_16\_sys & 16th percentile of total modeled equivalent width of CIV1548 absorption system (\AA) \\ \hline
Wr1548\_84\_sys & 84th percentile of total modeled equivalent width of CIV1548 absorption system (\AA) \\ \hline
Wr1550\_sys & Total modeled equivalent width of CIV1550 absorption system (\AA) \\ \hline
Wr1550\_16\_sys & 16th percentile of total modeled equivalent width of CIV1550 absorption system (\AA) \\ \hline
Wr1550\_84\_sys & 84th percentile of total modeled equivalent width of CIV1550 absorption system (\AA) \\ \hline
dv\_tau90 & Velocity width of CIV absorption system estimated from the interval of 5th to 95th percentile of cumulated optical depth ($\rm km~s^{-1}$) \\ \hline
dv\_lc & Velocity width of CIV absorption system estimated from the velocity difference between components with lowest and highest redshifts ($\rm km~s^{-1}$) \\ \hline
S/N\_abs & Signal-to-noise ratio of spectra surrounding the CIV absorption systems \\ \hline
Miss & Whether CIV absorption component is missed \\ \hline
False\_positive & Determining whether the CIV absorption component is a false-positive detection. a\_ions for type a and b\_type for type b false-positive detections \\ \hline
comment & Additional comments \\ \hline
\hline
\end{longtable*}

We also supply the simulated absorption line catalog as supplementary material, with the description of catalog columns shown in Table \ref{tab:simu_desc}. This catalog enables the readers to measure the sample completeness of the observed HIERACHY/MIKE \ion{C}{4} absorbers according to their own scientific goals.

\begin{table*}
\caption{Descriptive table for the simulated line catalog. The recorded lines are those fitted lines that either recover the parameters of the injected lines or exhibit the smallest velocity difference relative to them.}
\label{tab:simu_desc}
\centering
\renewcommand{\arraystretch}{1.3}
\begin{tabular}{ll}
\hline
\hline
Column Name & Description \\
\hline
QSO & Quasar name \\
z & The median redshift of the fitting box at $v = 0~\rm km~s^{-1}$ \\
SNR\_1548 & Signal-to-Noise-Ratio at CIV1548 \\
SNR\_1550 & Signal-to-Noise-Ratio at CIV1550 \\
logN\_i & Column density for the injected CIV absorption signal \\
b\_i & Doppler b-factor for the injected CIV absorption signal ($\rm km~s^{-1}$) \\
v\_i & Velocity center for the injected CIV absorption signal ($\rm km~s^{-1}$) \\
logN\_r & Column density for the recorded CIV absorption signal \\
b\_r & Doppler b-factor for the recorded CIV absorption signal ($\rm km~s^{-1}$) \\
v\_r & Velocity center for the recorded CIV absorption signal ($\rm km~s^{-1}$) \\
Wr\_i & Equivalent width for the injected CIV absorption signal (\AA) \\
Wr\_err\_i & Error in equivalent width for the injected CIV absorption signal (\AA) \\
Wr\_r & Equivalent width for the recorded CIV absorption signal (\AA) \\
Wr\_err\_r & Error in equivalent width for the recorded CIV absorption signal (\AA) \\
Chi\_before & Chi-squared before fitting \\
Chi\_after & Chi-squared after fitting \\
Frac\_contfwhm\_r & Fraction of pixels within |v| < 0.5v\_FWHM of recorded absorption signal are contaminated \\
Frac\_contfwhm2\_r & Fraction of pixels within |v| < v\_FWHM of recorded absorption signal are contaminated \\
Frac\_contfwhm\_i & Fraction of pixels within |v| < 0.5v\_FWHM of injected absorption signal are contaminated \\
Frac\_contfwhm2\_i & Fraction of pixels within |v| < v\_FWHM of injected absorption signal are contaminated \\
\hline
\end{tabular}
\end{table*}

\section{Summary}
\label{sec:Summary}

We conduct a comprehensive \ion{C}{4} blind survey at $z\approx 3-5$ using the high-SN and high-resolution Magellan/MIKE spectra of 25 quasars with $z_{\rm QSO} \approx 4-5$ obtained in the HIERACHY program.
The HIERACHY program is designed to study the \ion{He}{2} reionization and other related physical processes at the same redshift range.
In this paper, we report a catalog of 626 (1385) \ion{C}{4} absorption systems (components) 
with a limiting column density down to log$(N_{\rm CIV}/\rm cm^{-2}) \approx 12.3$ at the 50\% completeness (Fig. \ref{fig:complt}).
Among these systems, 557 are intervening absorption systems (1090 intervening components), which are used to study the statistics of \ion{C}{4} absorbers and associated cosmic evolution in \S\ref{sec:stat}.
The measured absorption properties are released in a machine-readable table introduced in \S\ref{sec:catalog}.

This HIERACHY/MIKE \ion{C}{4} sample significantly enlarges the number of \ion{C}{4} absorption systems at $z\approx 4$ (Panel a of Fig. \ref{fig:scomp}), where \ion{He}{2} in the IGM is expected to exhibit
significant ionization (Panel e of Fig. \ref{fig:comp_civ}). 
In particular, the high-quality MIKE spectra enable the detection of weak absorption systems with log($W_{\rm r,1548}/$\AA) $\approx -2.5$ to $-1$ or log$(N_{\rm CIV}/\rm cm^{-2}) \approx 12- 13$ (Panels d and e of Fig. \ref{fig:scomp}).

With the intervening HIERACHY/MIKE sample, we examine the cosmological evolution of \ion{C}{4} absorbers.
First, we detect a significant breakpoint at $\approx 13.2$ in the \ion{C}{4} CDDF at $z\approx 3-5$, suggesting a broken power-law fitting (Fig. \ref{fig:cddf}).
In particular, the turn-over column density is log$(N_{\rm crit}/\rm cm^{-2}) = 13.35^{+0.20}_{-0.19}$, with the measured slopes of $\alpha = -1.04^{+0.20}_{-0.26}$ below the critical column density and $\beta = -2.16^{+0.16}_{-0.14}$ above.
By dividing the full sample into two redshift bins of $z\approx3-4$ and $z\approx 4-5$, we found that all CDDF parameters are consistent within 0.5 $\sigma$.
Although the HIERACHY/MIKE \ion{C}{4} sample is already the largest at $z\approx 3-5$, the study of cosmological evolution of the CDDF is still limited by the sample size.
Besides the MIKE spectra, we currently also obtained the Magellan/MagE spectra of $\approx 50$ QSOs at $z\approx 4.6-5.1$ with a relatively lower spectral resolution of $R\approx 7000$ to better constrain the evolution of absorbers with $\log N_{\rm CIV}/{\rm cm^{-2}}\gtrsim 13$.
The observation information of 29 reduced MagE QSO spectra can be found in \citetalias{Li24}, while the \ion{C}{4} catalog will be reported in a future paper.

In addition to the CDDF, we also measure the comoving path length number density $dn_{\rm CIV}/dX$ and the cosmic mass abundance $\Omega_{\rm CIV}$ for the intervening \ion{C}{4} absorption systems with log$(N_{\rm CIV}/\rm cm^{-2}) \gtrsim 13.2$.
Both tracers of the cosmological evolution of \ion{C}{4} show an increase ($\approx$ 2.2 and 1.4 $\sigma $) as the redshift decreases (Fig. \ref{fig:evol_com}).
The evolution of measured $\Omega_{\rm CIV}$ is consistent with previous studies, suggesting a shallower increase of \ion{C}{4} cosmic mass abundance at $z\lesssim5$ (e.g., \citealt{Simcoe11}; \citetalias{Hasan20, Davies23b}), 
which are typically based on results bridging different samples at different redshift ranges, or on the limited sample size at $z\approx3-5$.
Different from high-column-density \ion{C}{4} absorbers, the abundance of weak \ion{C}{4} systems with log$(N_{\rm CIV}/\rm cm^{-2}) \approx 12.3-13.2$ exhibit a constant $dn_{\rm CIV}/dX$ and $\Omega_{\rm CIV}$ at $z\approx 3-5$.
Therefore, we conclude that the cosmic evolution of \ion{C}{4} is driven by \ion{C}{4} absorbers with log$(N_{\rm CIV}/\rm cm^{-2}) \gtrsim 13.2$.

Distinguishing whether metal enrichment from galactic feedback or UVB photoionization is the primary driver for the observed evolution of \ion{C}{4} requires additional constraints from other ion absorptions, as these processes have distinct effects on the evolution of metal ion abundances (\citealt{Finlator15}). While metal enrichment boosts both populations, photoionization increases highly-ionized ion abundance and reduces lower-ionized ion abundance. This theoretical understanding helps explain the observed transition from weak or undetectable CIV absorption in low-ionization systems at $z\approx6$ to their commonness at $z\approx3$ (e.g., \citealt{Rubin15,Becker19,Cooper19}), which may result from increases in the hardness or amplitude of the UVB following HI reionization. The HIERACHY/MIKE and HIERACHY/MagE samples are particularly valuable for studying this transition, as they currently provide the largest collection of high- and medium-resolution quasar spectra spanning the critical redshift range of $z\approx4-5$.

\section*{Acknowledgements}

The authors would like to acknowledge Prof. Ian U. Roederer from the North Carolina State University for his contributions in observations and data analysis.
We also acknowledge Farhanul Hasan and Rebecca L. Davies for sharing their \ion{C}{4} catalog and data.
We thank the anonymous referee for their thorough review and constructive comments, which have significantly improved the quality of this manuscript.
CZ acknowledge support from the National Key R\&D Program of China (grant no.~2023YFA1605600), the National Science Foundation of China (grant no.~12073014), the science research grants from the China Manned Space Project with No.~CMS-CSST-2021-A05, and Tsinghua University Initiative Scientific Research Program (No.~20223080023).
J.T.L. acknowledges the financial support from the National Science Foundation of China (NSFC) through the grants 12273111 and 12321003, and also the science research grants from the China Manned Space Project. X. W. is supported by the National Natural Science Foundation of China (grant 12373009), the CAS Project for Young Scientists in Basic Research Grant No. YSBR-062, the Fundamental Research Funds for the Central Universities, the Xiaomi Young Talents Program, and the science research grant from the China Manned Space Project.

\bibliographystyle{mnras}
\bibliography{HIERACHY_III}

\appendix

\section{Intervening regions of HIERACHY/MIKE \ion{C}{4} sample}\label{sec:spec}

\begin{figure*}
\begin{center}
\includegraphics[width=1\textwidth]{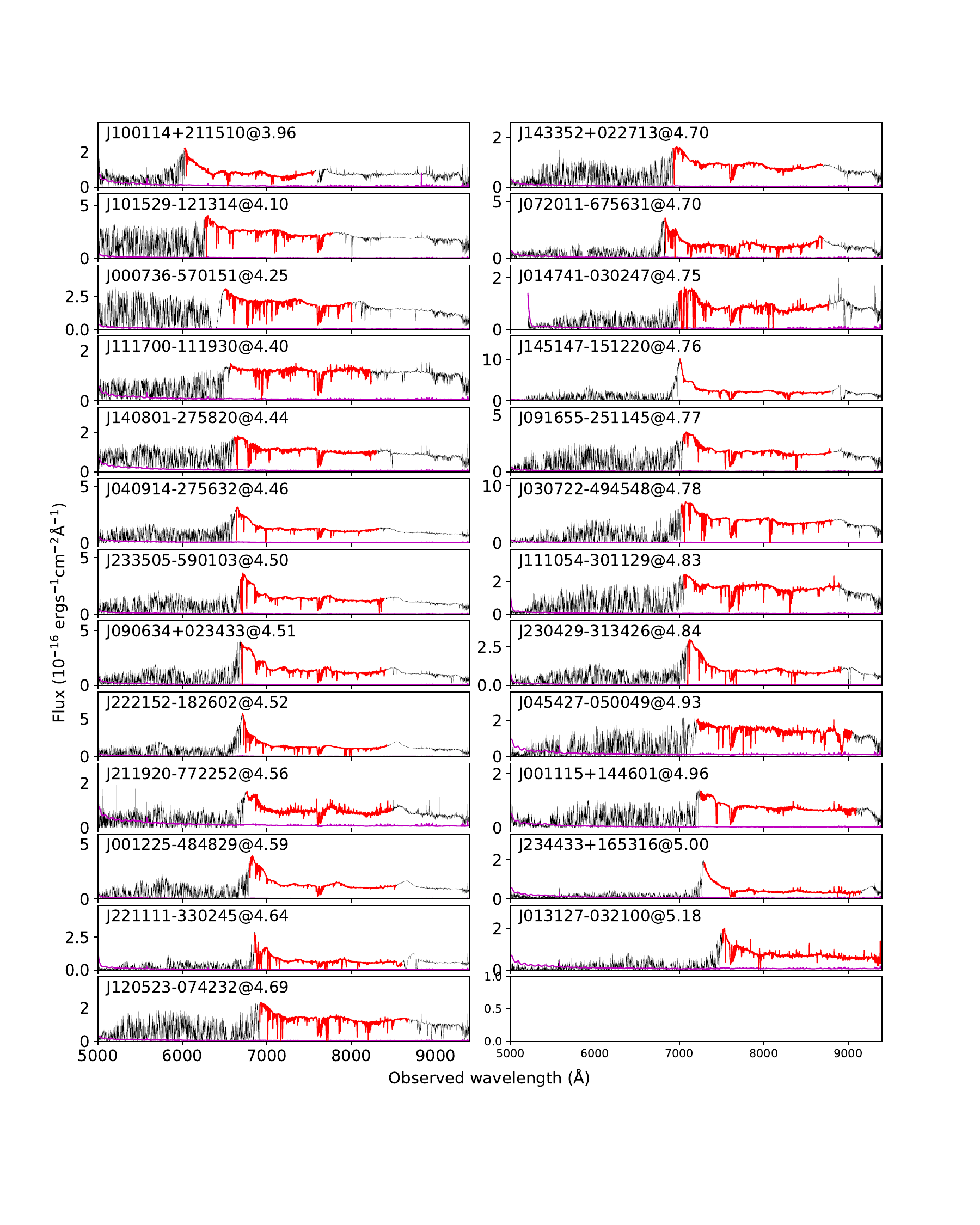}
\vspace{-0.75in}
\caption{A schematic view of 25 smoothed quasar spectra from the red channel of MIKE (black curves). The magenta and red curves show the $1\sigma$ flux error and the intervening \ion{C}{4} search region, respectively. The intervening region is typically defined as the wavelength region from Ly$\alpha$ emission line to the region with blueshift velocity $\Delta v_{\rm b} > 5000\rm~km~s^{-1}$ relative to quasar redshift. Our \ion{C}{4} absorption line catalog includes all absorption lines identified in the wavelength region redward of quasar Ly$\alpha$ emission lines. However, we only use the absorption lines within the intervening regions to study the statistical properties of \ion{C}{4} absorber (\S\ref{sec:stat}), to alleviate the contamination of quasar intrinsic absorption.} 
\label{fig:spectra}
\end{center}
\end{figure*}

We show the intervening \ion{C}{4} search regions with red curves for 25 HIERACHY/MIKE quasar spectra in Fig. \ref{fig:spectra}.
In particular, to avoid the quasar intrinsic absorption, we set the maximum wavelengths of these regions based on a blueshift velocity of $\Delta v_{\rm b} = 5,000\rm~km~s^{-1}$ relative to quasar redshift $z_{\rm QSO}$. 
The minimum wavelengths are generally set by the $\lambda_{\rm l} = \lambda_{\rm Ly\alpha}\times (1+z_{\rm QSO})$.
If the derived $\lambda_{\rm l}$ significantly deviate from the peak of broad emission feature (Ly$\alpha$ emission line in most cases), where their blue wings are absorbed by Ly$\alpha$ forests, we manually allocate the $\lambda_{\rm l}$ to this peak.
For example, with the QSO redshift $z=4.25$ reported in \citealt{Wolf20}, the calculated $\lambda_{\rm l}$ for QSO J000736-570151 is 6382~\AA, which lies within a damped Ly$\alpha$ absorption line.
Although we have searched for \ion{C}{4} absorption lines down to a wavelength of 6382~\AA,
We still manually adjust $\lambda_{\rm l}$ of intervening region to the peak of the broad emission feature at approximately 6500~\AA~(Fig. \ref{fig:spectra}),
to avoid the effects of continuum placement uncertainties on \ion{C}{4} detection.

\section{Contamination}\label{sec:cont}
  
Searching for \ion{C}{4} absorption lines is often affected by contaminants. 
These contaminants can change the strength (or velocity) ratio between doublet lines.
This ratio is a basic tracer for identifying \ion{C}{4} $\lambda\lambda 1548.2, 1550.8$ doublets.
To increase the detection rate and reliability, we define several contamination types and use them in the automatic search for \ion{C}{4} absorption lines, inspection of identified lines, and assignment of line confidence.
We summarize the description, definition, and usage of these contamination types in Table \ref{tab:cont}.

In this paper, we consider contaminated pixels $\rm {C_{p}}$ those for which the \ion{C}{4} doublet shows an excess data absorption ($F$) relative to the modeled \ion{C}{4} absorption profile ($M_{\rm CIV}$), i.e., $|(d\chi(i-1)+d\chi(i)+d\chi(i+1))/\sqrt{3}| > 3$, where $d\chi(i)=\frac{F(i)_{1548}-M_{\text CIV}(i)_{1548}}{E(i)_{1548}} - \frac{F(i)_{1550}-M_{\text CIV}(i)_{1550}}{E(i)_{1550}}$. Contaminated pixels $\rm {C_{p}}$ are in \ion{C}{4} $\lambda1548.2$ and $\lambda1550.8$ absorption profile under the conditions $d\chi <-3~\text{and}~> 3$, respectively. 

In practice, telluric absorption contamination $\rm {C_{p,tel}}$ could be well modeled by current telluric transmission template with a parameter $\alpha_{\rm T}$ to adjust the absorption strength $M_{\rm tel}(i) = T^{\alpha_{\rm T}}(i)$ (e.g., \citealt{Bertaux14} as shown in Panel d of Fig. \ref{fig:lv}).
Therefore, in the line search step, we generally search for contamination pixels $\rm {C_{p,oth}}$ arising from unwell-modeled contamination (contamination other than telluric) but not $\rm {C_{p}}$.
The $\rm {C_{p,oth}}$ can be defined as an excess absorption in data profile relative to modeled absorption profile $M$, where $M$ equals $M_{\rm tel}$ before adding any \ion{C}{4} component and becomes $M_{\rm CIV} \times M_{\rm tel}$ after the \ion{C}{4} components are added.
This excess absorption is searched and masked before identifying a \ion{C}{4} absorption component (i.e., adding a Voigt profile).

Inspecting the contamination sources for these unwell-modeled contamination pixels $\rm {C_{p,oth}}$ is an important part of line inspection. 
In this paper, we generally inspect three kinds of contamination sources for $\rm {C_{p,oth}}$. (1) The relic sky emission lines $\rm {C_{p,oth-sky}}$ in the wavelength regions where $F_{\rm sky}/C_{\rm sky} > 3$. $F_{\rm sky}$ is the flux of sky background spectra produced in sky subtraction procedure and $C_{\rm sky}$ is the `sky background continuum' produced with global-continuum estimation method described in \citetalias{Li24}. We show an example of relic sky emission in the Panel c of Fig. \ref{fig:lv}. (2) The superposition of an adjacent \ion{C}{4} 1548.2 (1550.8) absorption line (denoted as $\rm {C_{p,oth-CIV}}$) on searched \ion{C}{4} 1550.8 (1548.2) absorption lines, as illustrated in the Panel e of Fig. \ref{fig:lv}. This contamination could be alleviated by enlarging line search (fitting) box. However, this is at the cost of increasing the probability of misidentifying weak absorption lines. (3) The superposition of absorption features from other ions $\rm {C_{p,oth-ion}}$ at other redshifts, as shown in the Panel f of Fig. \ref{fig:lv}. In this paper, we inspect contamination from ions that are commonly observed in quasar spectra (Table \ref{tab:cont}). We perform the automatic line search for these ions and inspect whether the contamination in the \ion{C}{4} absorption profile is superimposed by their modeled absorption profiles. The automatic line search method introduced in \citetalias{Li24} is not valid for single lines such as \ion{C}{2}1334, \ion{O}{1}1302 and \ion{Al}{2}1670. We therefore artificially produce a `doublet' \ion{C}{2}1334 and \ion{O}{1}1302, i.e., set their oscillator strength ratio to one and consider the deviation in absorption strength and width as contamination. This is motivated by observed \ion{C}{2}1334 and \ion{O}{1}1302 generally have similar equivalent width \citep{Becker19,Sebastian24}. We then inspect other ions listed in Table \ref{tab:cont} and consider a positive detection as the `doublet' is aligned by at least one of these ions at redshift space. We identify approximately 20 \ion{C}{2}1334 absorption systems paired with other ions, a number consistent with the $dn_{\rm CII}/dX$ measured at $z\approx5$ \citep{Sebastian24}. This suggests that the current sample of \ion{C}{2}1334 and these single lines should be complete enough for the goal of contamination inspection. 

\begin{table*}
\caption{Contamination types used in this paper.}
\centering
\begin{tabular}{cccc}
\hline
\textbf{Symbol} & \textbf{Description} & \textbf{Criteria/Definition}  & \textbf{Comments} \\
\hline
$\rm {C_{p}}$ & \parbox{1.5in}{\centering excess absorption in data profile relative to modeled \ion{C}{4} profile} & \parbox{2.5in}{\centering$|(d\chi(i-1)+d\chi(i)+d\chi(i+1))/\sqrt{3}| > 3$\\ $d\chi(i)=\frac{F(i)_{1548}-M_{\rm CIV}(i)_{1548}}{E(i)_{1548}} - \frac{F(i)_{1550}-M_{\rm CIV}(i)_{1550}}{E(i)_{1550}}$} & \parbox{2in}{\centering used in $\rm Lv_{\rm fit}$ assignment for observed and simulated lines} \\
\hline
$\rm {C_{p,tel}}$ & telluric contaminated pixels & $M_{\rm tel}(i) = T_{\text{TAPAS}}^{\alpha_{\text T}(i)}$ &\parbox{2in}{\centering basic model component in the fit and MCMC}\\
\hline
$\rm {C_{p,oth}}$ & contamination pixels other than telluric& \parbox{2.5in}{\centering$|(d\chi(i-1)+d\chi(i)+d\chi(i+1))/\sqrt{3}| > 3$\\ $d\chi(i)=\frac{F(i)_{1548}-M(i)_{1548}}{E(i)_{1548}} - \frac{F(i)_{1550}-M(i)_{1550}}{E(i)_{1550}}$} & \parbox{2in}{\centering used as the mask for contamination not well modeled in line search and simulation step} \\
\hline
$\rm {C_{p,oth-sky}}$ & sky contamination in $\rm {C_{p,oth}}$ & $F_{\text{sky}}(i)/C_{\text{sky}}(i) > 3$ & \parbox{2in}{\centering used in visual inspection} \\
\hline
$\rm {C_{p,oth-CIV}}$ & adjacent CIV contamination in $\rm {C_{p,oth}}$ & \parbox{2.5in}{\centering another CIV1550 (CIV1548) absorption profiles superimpose on searched CIV1548 (CIV1550) profiles} & \parbox{2in}{\centering used in visual inspection} \\
\hline
$\rm {C_{p,oth-ion}}$ & other ion contamination in $\rm {C_{p,oth}}$ & \parbox{2.5in}{\centering other ion absorption profiles at other redshifts superimpose on searched CIV1548 (CIV1550) profiles} & \parbox{2in}{\centering used in visual inspection, ions include \ion{Si}{4}1393,1403; \ion{N}{5}1239,1243; \ion{Mg}{2}2796,2804; \ion{Fe}{2}2344,2382,2587,2600; \ion{C}{2}1334; \ion{O}{1}1302; \ion{Si}{2}1260,1304,1526; \ion{Al}{2}1670}. \\
\hline
\end{tabular}
\label{tab:cont}
\end{table*}

\section{Confidence level}\label{sec:conf}

The $\rm Lv_{MCMC}$ assignment in the inspection step generally follows the $\rm Lv_{fit}$ assignment in the automatic line search steps, which broadly classifies searched lines with line significance and contamination level. 
This process introduces two key refinements: First, we further divide the absorption lines with high-level contamination into two subgroups: contamination with known and unknown sources. Second, we add a confidence level for potential false-positive detections.
We herein show some examples of $\rm Lv_{MCMC}$ assignment for inspected lines.

\begin{figure*}
\begin{center}
\includegraphics[width=0.49\textwidth]
{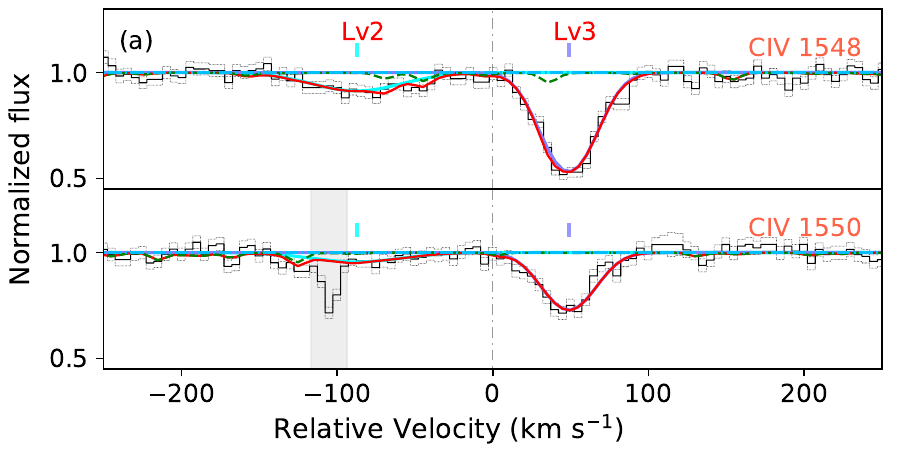}
\includegraphics[width=0.49\textwidth]
{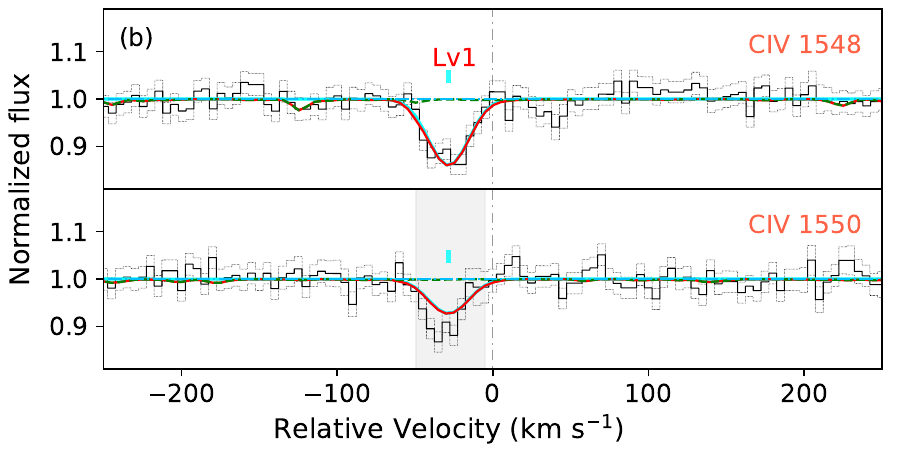}
\includegraphics[width=0.49\textwidth]
{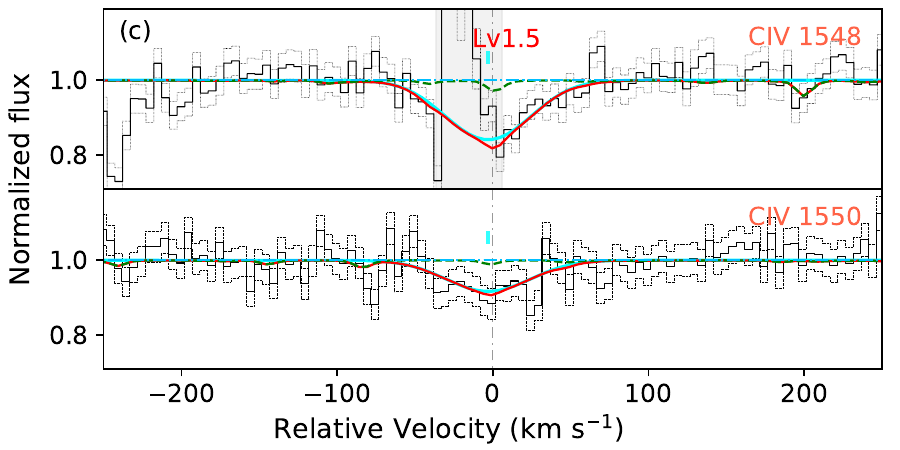}
\includegraphics[width=0.49\textwidth]
{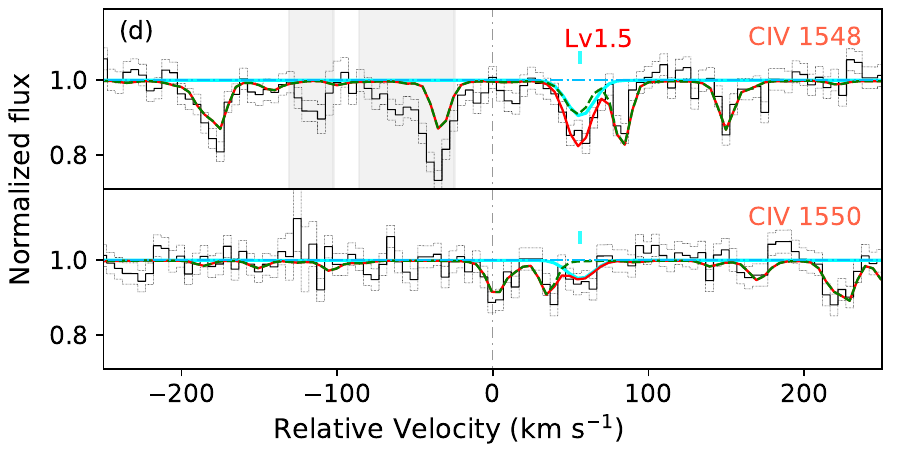}
\includegraphics[width=0.49\textwidth]
{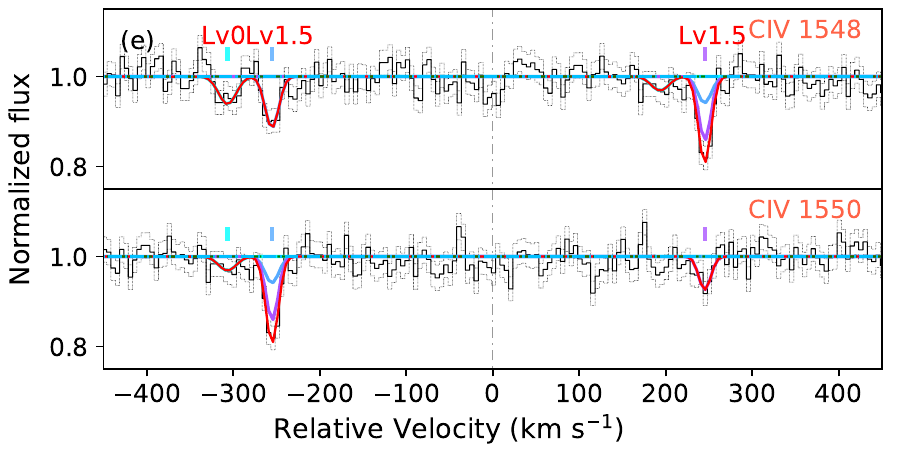}
\includegraphics[width=0.49\textwidth]
{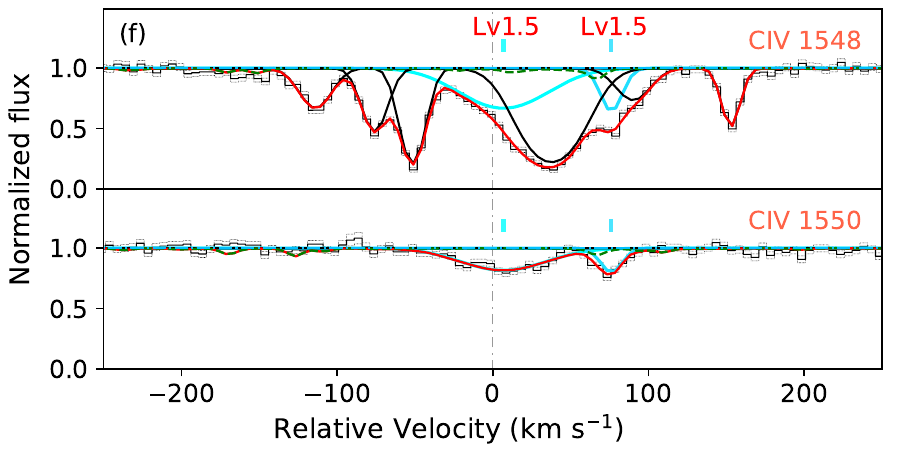}
\includegraphics[width=0.49\textwidth]
{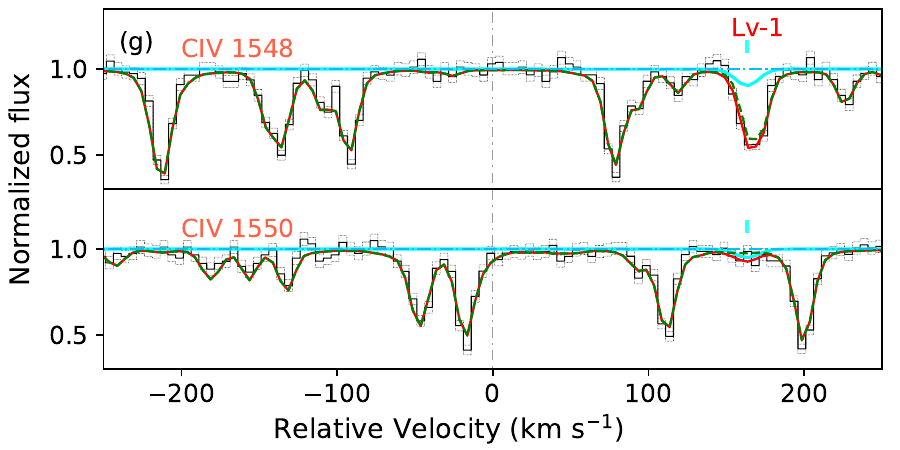}
\includegraphics[width=0.49\textwidth]
{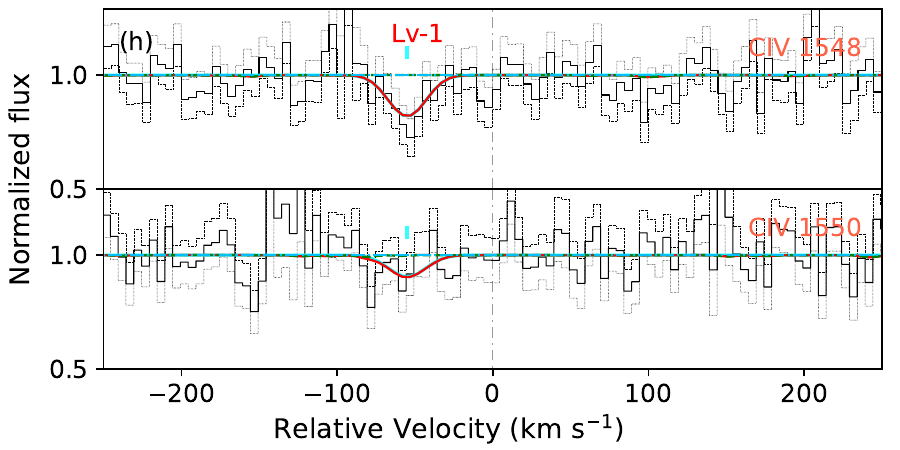}
\caption{Examples of $\rm Lv_{MCMC}$ assignment for inspected \ion{C}{4} absorption lines. We show the $\rm Lv_{MCMC}$ of each inspected \ion{C}{4} absorption line with red labels, which are located above colored vertical lines representing their velocity positions. Black step curves represent the normalized data flux with 1$\sigma$ error. Red lines show the total model profile, which generally involves \ion{C}{4} Voigt absorption (colored curves), telluric absorption (green dashed curves) and occasionally involves other ion absorption (black curves in Panel f). The shaded regions show the masked contamination features.} 
\label{fig:lv}
\end{center}
\end{figure*}

We show examples of low-significance absorption lines with $\rm Lv_{MCMC} = 0$ in Panels e of Fig. \ref{fig:lv}. These absorption lines have measured $N/\Delta N < 3$ in the MCMC fitting, where $\Delta N$ is calculated using the interval between the 16\% and 50\% percentiles of MCMC chains.
We do not use them in the construction of absorption systems and statistical studies in \S\ref{sec:stat},
as they have a high probability of being produced by white noise, and the identification of these absorption lines is not robust.

In this paper, the identification of high-significance absorption lines with $\rm Lv_{MCMC} = 3$ and 2 generally has the highest robustness.
Absorption lines with $\rm Lv_{MCMC} = 3$ are absorption features without contamination pixels $\rm {C_{p}}$. 
We show an example of $\rm Lv_{MCMC} = 3$ absorption line in Panel a of Fig. \ref{fig:lv}.
The low-level contaminated absorption lines $\rm Lv_{MCMC} = 2$ have $\rm {C_{p}}$ within their absorption profiles, however these $\rm {C_{p}}$ only occupy a small fraction of pixels ($<0.5$) within the core of absorption profiles, defined as $|v| < 0.5v_{\rm FWHM}$.
These contaminated pixels generally do not affect the identification of \ion{C}{4} absorption lines, as shown in Panel a.

The estimation of identification robustness for absorption lines with high-level contamination is more complex.
These absorption lines have more than half of the pixels within $|v| < 0.5v_{\text{FWHM}}$ of the modeled \ion{C}{4} line profile classified as $\rm {C_{p}}$.
To enhance the identification robustness of these lines, we further inspect four kinds of contamination sources for them: Panel c for relic sky emission line, Panel d for telluric absorption, Panel e for adjacent \ion{C}{4} absorption, and Panel f for other ion absorption at other redshifts. 
We assign $\rm Lv_{MCMC} = 1.5$ to a absorption line if one of the above contamination source could explain $\rm {C_{p}}$. If none of these contamination sources can explain the high-level contamination, we follow the confidence level assigned in automatic line search step and assign these absorption lines with $\rm Lv_{MCMC} = 1$, as shown in Panel b.

We further make the following minor adjustments to the confidence level assignment, as above general rules cannot correctly address the following special cases.
A significant absorption line within strongly blending absorption systems can have $N/\Delta N < 3$. Additionally, a broad absorption system may have its central absorption component contaminated by unknown sources. We generally assign them with $\rm Lv_{MCMC} = 3$ and 2 rather than $\rm Lv_{MCMC} = 0$ and 1, to avoid breaking an absorption system with continuous absorption profile.
Moreover, we typically ignore the contamination level and assign a $\rm Lv_{MCMC} = 3$ or 2 to \ion{C}{4} absorption lines that are associated with other ion absorptions at $z_{\rm ion}\approx z_{\rm CIV}$. In this paper, associated absorption lines are defined as ion absorption features within the \ion{C}{4} absorption profiles (in redshift space). 
We do not require them to have the same redshift, as we observe redshift deviations in some apparent associated cases.

After assigning the $\rm Lv_{MCMC}$, we then inspect for the false-positive detections for all $\rm Lv_{MCMC}$ levels and assign $\rm Lv_{MCMC} = -1$ to them.
We generally search for two types of false-positive detections: ion absorption other than \ion{C}{4} doublet (type a) and profile mismatches between doublet components (type b).
Overall, we identify 55 type a and 43 type b false-positive detections among absorption components with $N/\Delta N > 3$.

During the false-positive detection inspection, we generally exclude \ion{C}{4} absorption components that show associated \ion{Si}{4} absorption in redshift space. This exclusion is based on the extremely low probability of four unrelated lines randomly aligning to form two associated doublets at the same redshift.
All inspections are conducted within detectable regions, i.e., wavelength region redward of Ly$\alpha$ forest and redshift region smaller than quasar redshift.
We use all ions listed in Table \ref{tab:cont}.
For \ion{C}{2}1334, \ion{O}{1}1302, \ion{Si}{2}1260,1304,1526 and \ion{Al}{2}1670, we only inspect wavelength regions associated with $\approx20$ identified \ion{C}{2}1334 systems described in Appendix \ref{sec:cont}.
For \ion{Fe}{2}2344,2382,2587,2600, we typically require detection of all four lines or associated \ion{Mg}{2}2796,2804 absorption when detectable, as \ion{Mg}{2} absorption is generally stronger than \ion{Fe}{2} absorption.
Similarly, for \ion{Si}{4}1393,1403, we require an associated \ion{C}{4} if they are detectable.
Lastly, for ions often without stronger (\ion{Mg}{2}2796,2804) or apparent (\ion{N}{5}1239,1243) associated absorption, we primarily assess the goodness of fit and level of contamination.
Type b false-positive detections, shown in Panels g and h of Fig. \ref{fig:lv}, occur when doublet components exhibit profile mismatches, typically due to significant contamination or noise effects.

\section{Bias from Component Completeness}\label{sec:bias_comp}

Measuring the completeness of the absorption system is difficult because it is sensitive to the strength ratio and velocity separation of absorption components, which are poorly constrained in the observed system sample.
Therefore, in \S\ref{sec:stat}, we use the constructed component completeness from \S\ref{sec:complt} for observed absorption systmes.
This could introduce a bias that leads to an overestimation of absorber completeness in specific column density bins.

\begin{figure*}
\begin{center}
\includegraphics[width=0.99\textwidth]
{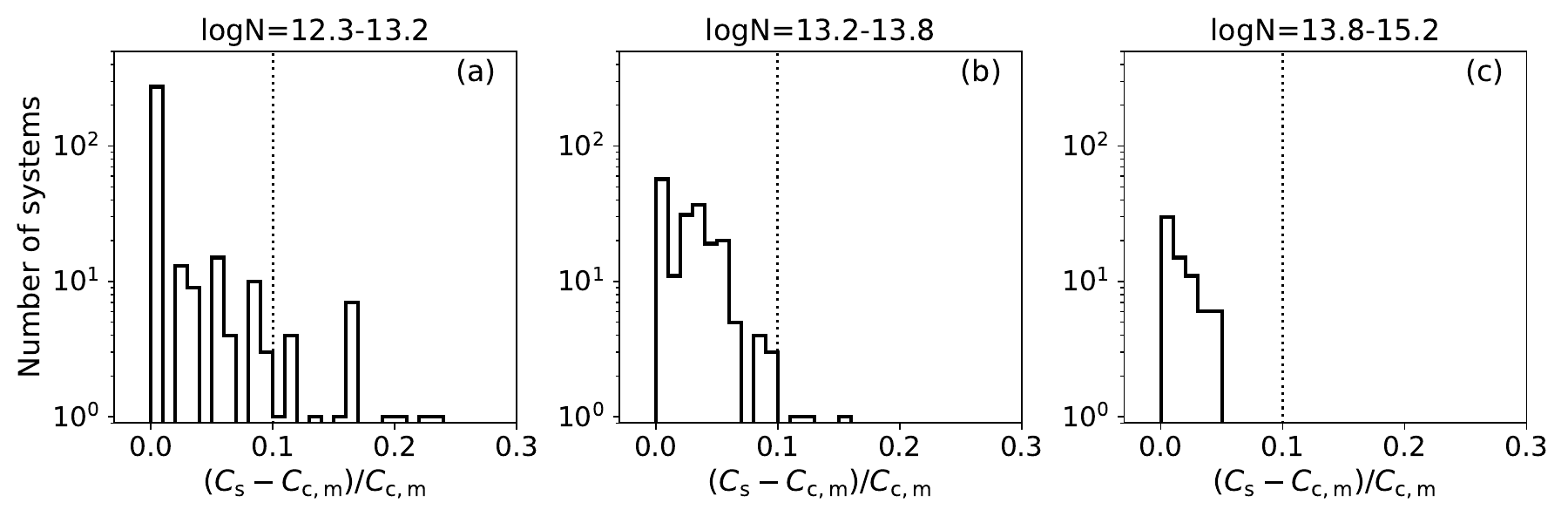}
\caption{Distribution of completeness bias values $(C_{\rm s}-C_{\rm c,m})/C_{\rm c,m}$ (defined in Appendix \ref{sec:bias_comp}) for HIERACHY/MIKE \ion{C}{4} absorption systems. The typical magnitudes of relative statistical uncertainties of CDDF, $dn_{\rm CIV}/dX$, and $\Omega_{\rm CIV}$ for these systems are represented by vertical dotted lines. We exclude an absorption system with $(C_{\rm s}-C_{\rm c,m})/C_{\rm c,m} \approx 0.7$ in the Panel a for clarity.}
\label{fig:bias}
\end{center}
\end{figure*}

To quantify this potential bias, we make the following assumption: the detectability of an absorption system is primarily determined by its strongest component. This assumption holds because only non-single-component systems  near the detection limit ($\sim 1\%$ for HIERACHY/MIKE \ion{C}{4} sample) or those containing numerous components with equal strength (also rare) would exhibit significantly different detectability between the system and its strongest component. 
With the above assumption, the interested bias manifests as an error in the estimated component completeness for an absorber when the system's strength is used to approximate that of its strongest component.
We quantify this error as $(C_{\rm s}-C_{\rm c,m})/C_{\rm c,m}$, where $C_{\rm s}$ and $C_{\rm c,m}$ are the completeness values for absorbers with their strength represented by the column density of system and strongest component, respectively. As shown in Fig. \ref{fig:bias}, most of Magellan/MIKE \ion{C}{4} absorbers have $(C_{\rm s}-C_{\rm c,m})/C_{\rm c,m}$ values smaller than 0.1, which is the typical magnitudes of relative statistical uncertainties reported in Tables \ref{tab:cddf}, \ref{tab:dnomegadx}, \ref{tab:cddf_comp}, and \ref{tab:dnomegadx_comp}.

\section{Absorption component statistics}\label{sec:stat_comp}

\begin{figure*}
\begin{center}
\includegraphics[width=0.49\textwidth]
{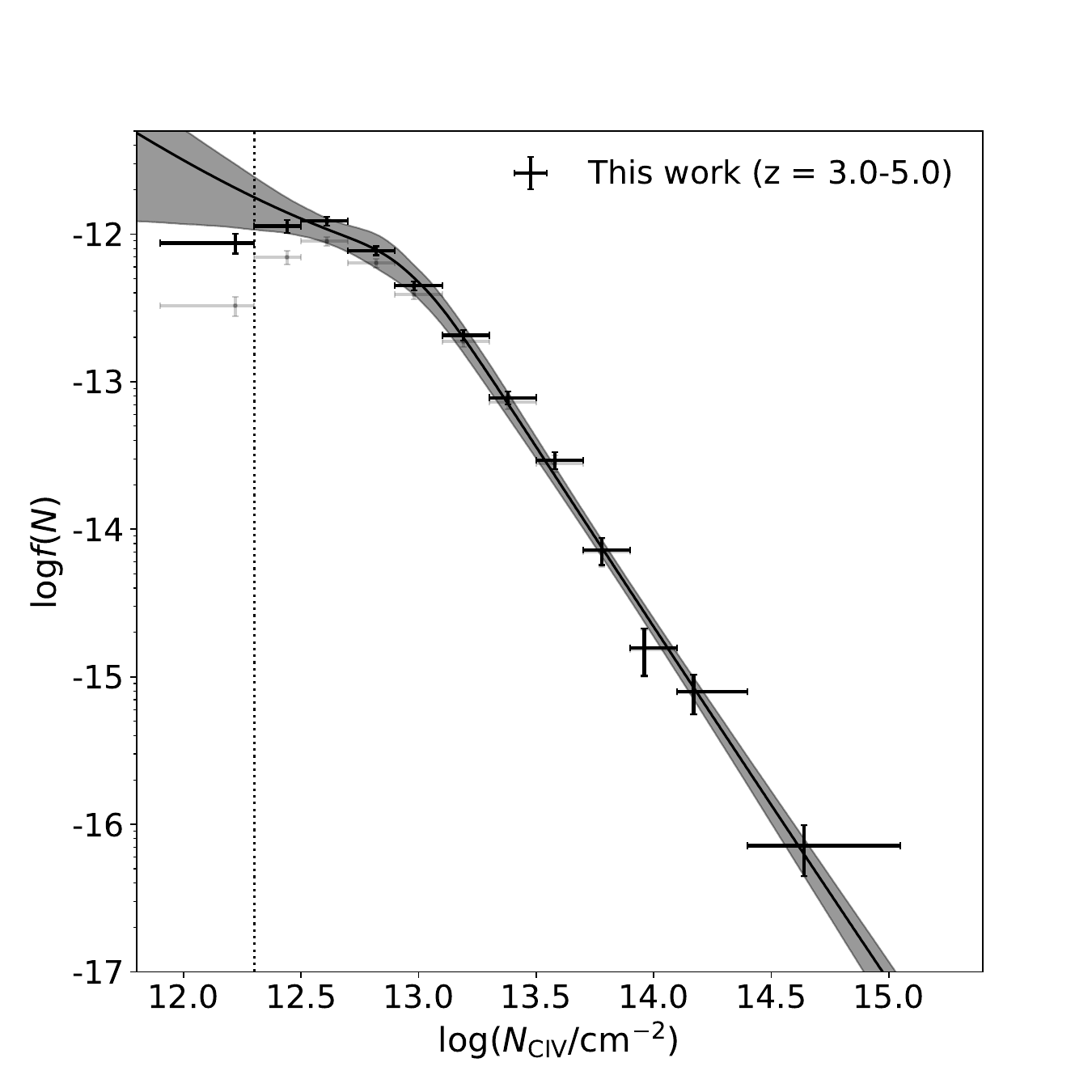}
\includegraphics[width=0.49\textwidth]
{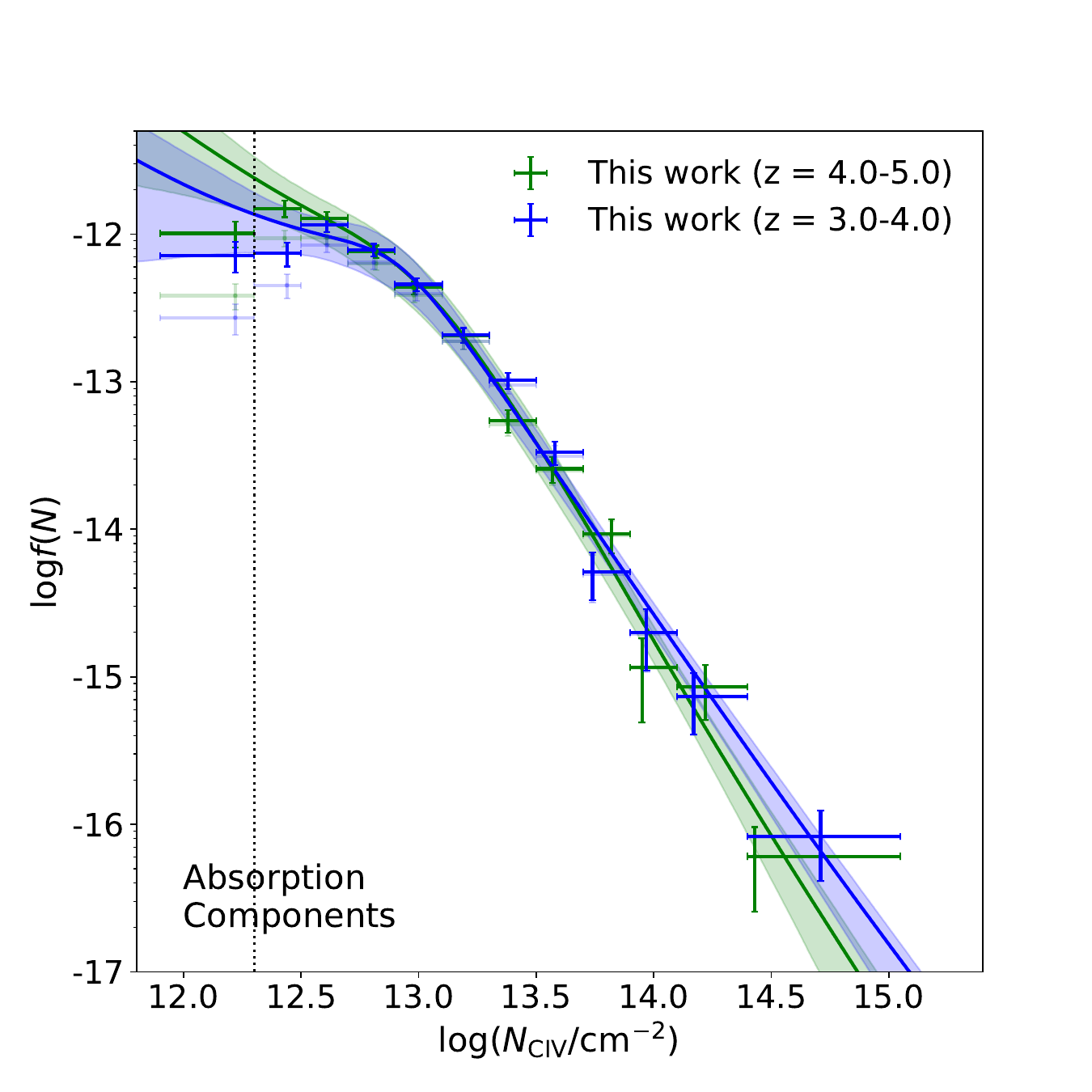}
\includegraphics[width=0.49\textwidth]
{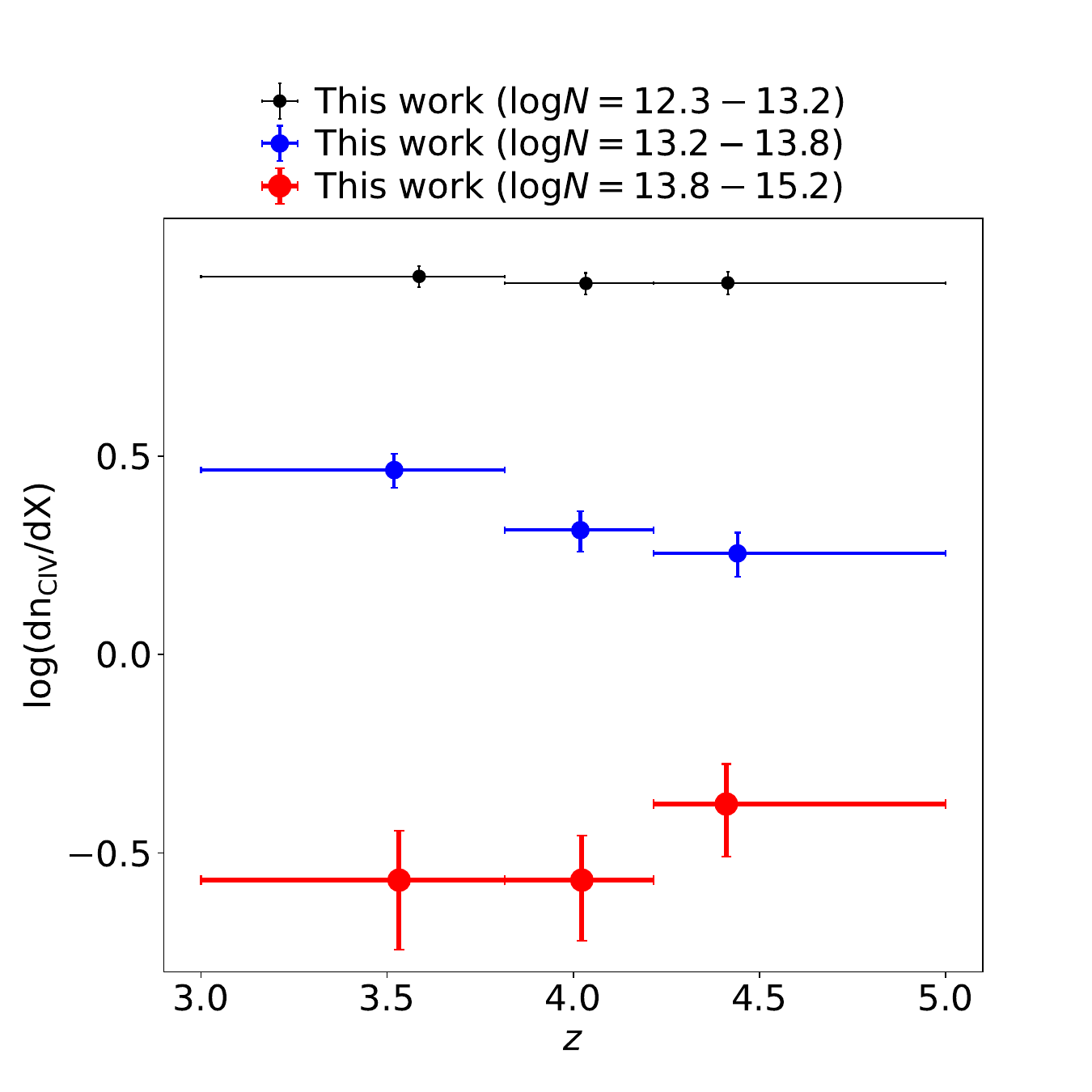}
\includegraphics[width=0.49\textwidth]
{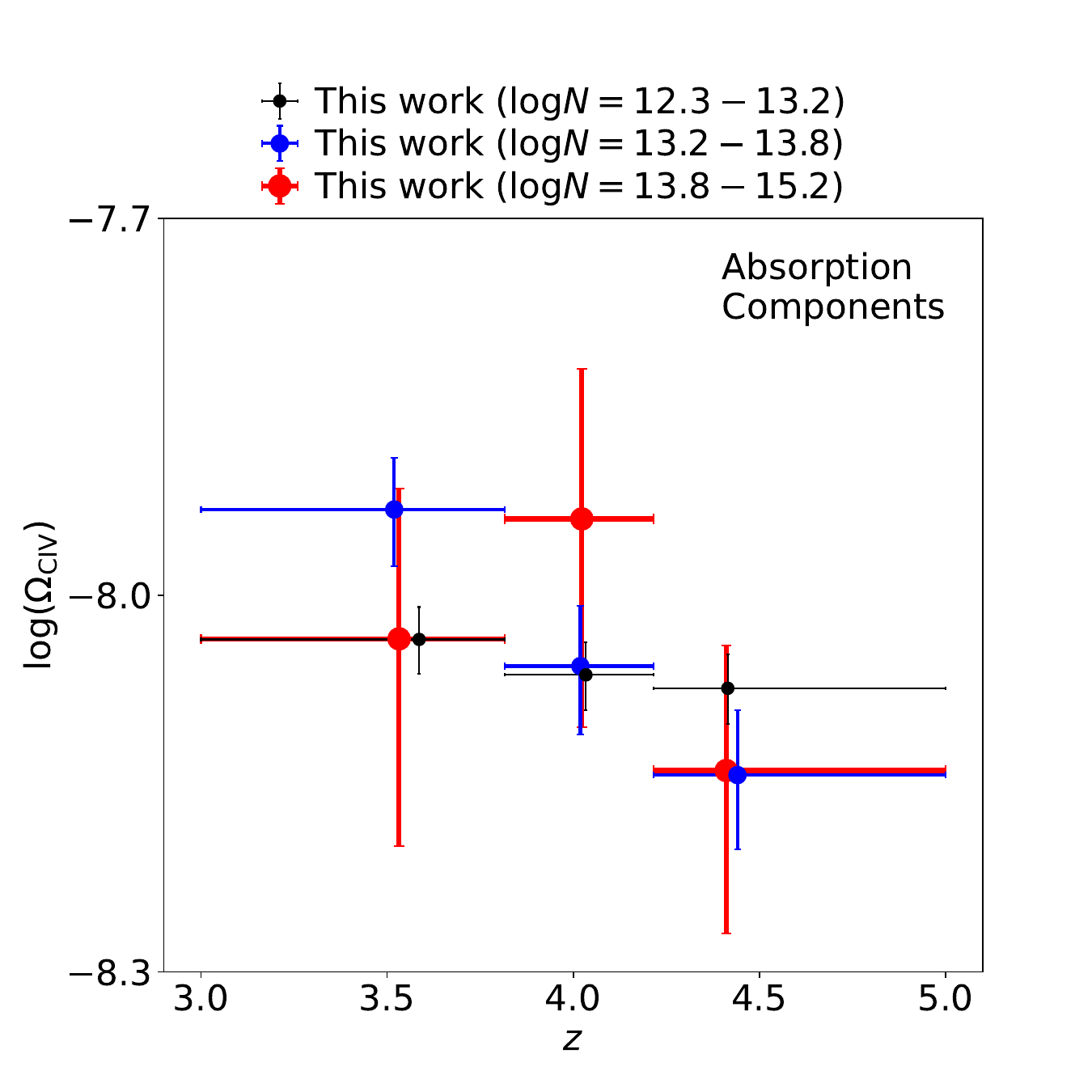}
\caption{The CDDF (upper panel), $dn_{\rm CIV}/dX$ (lower-left panel) and $\Omega_{\rm CIV}$ (lower-right panel) of the HIERACHY/MIKE \ion{C}{4} absorption components.} 
\label{fig:stat_comp}
\end{center}
\end{figure*}

Since system statistics may be more sensitive to cosmic abundance evolution than component statistics, we report system statistics in \S\ref{sec:stat} and perform similar measurements for absorption components in this section. The component CDDF (Table \ref{tab:cddf_comp}) exhibits a less conclusive broken power-law shape compared to the system CDDF, primarily because the measured critical column density (Table \ref{tab:tb_cddf_fit_comp}) is closer to the column density at 50\% completeness (upper Panel of Fig. \ref{fig:stat_comp}). For absorption components with log$(N_{\rm C IV}/\rm cm^{-2}) =13.8-15.2$, the measured $dn_{\rm CIV}/dX$ and $\Omega_{\rm CIV}$ do not show an increase as redshift decreases (lower Panel of Fig. \ref{fig:stat_comp} and Table \ref{tab:dnomegadx_comp}), in contrast to the trend observed in absorption systems (Fig. \ref{fig:evol_nbin}). These differences can be explained by the fact that component statistics typically decompose stronger absorber systems into multiple weaker absorption components, which tends to increase the abundance of weak absorbers while decreasing the abundance of strong absorbers (significant differences in the measured $\mathscr{N}$ values between Table \ref{tab:cddf} and Table \ref{tab:cddf_comp}, as well as between Table \ref{tab:dnomegadx} and Table \ref{tab:dnomegadx_comp}).

\begin{table*}
\caption{The measured CDDF for the HIERACHY/MIKE \ion{C}{4} absorption components in different redshift ranges.}
\label{tab:cddf_comp}
\centering
\begin{tabular}{cccccccccccc}
\hline\hline
 & \multicolumn{3}{c}{$ 3.0 < z < 5.0 $} & & \multicolumn{3}{c}{$ 3.0 < z < 4.0 $} & & \multicolumn{3}{c}{$ 4.0 < z < 5.0 $} \\ 
 \cline{2-4} \cline{6-8} \cline{10-12}
 log$N$ range & log$N_{\rm Med}$& $\mathscr{N}$ & log$f(N)$  & &log$N_{\rm Med}$& $\mathscr{N}$ & log$f(N)$  & &log$N_{\rm Med}$&  $\mathscr{N}$ & log$f(N)$ \\ 
\hline
11.9 - 12.3 & 12.22 & 44 & $-12.06 \pm 0.07$ & & 12.22 & 18 & $-12.15 \pm 0.10$ & & 12.22 & 26 & $-11.99 \pm 0.09$ \\
12.3 - 12.5 & 12.44 & 91 & $-11.95 \pm 0.05$ & & 12.44 & 29 & $-12.13 \pm 0.08$ & & 12.43 & 62 & $-11.83 \pm 0.06$ \\
12.5 - 12.7 & 12.61 & 185 & $-11.91 \pm 0.03$ & & 12.61 & 86 & $-11.94 \pm 0.05$ & & 12.61 & 99 & $-11.89 \pm 0.04$ \\
12.7 - 12.9 & 12.82 & 209 & $-12.11 \pm 0.03$ & & 12.81 & 104 & $-12.11 \pm 0.04$ & & 12.82 & 105 & $-12.12 \pm 0.04$ \\
12.9 - 13.1 & 12.98 & 202 & $-12.35 \pm 0.03$ & & 12.99 & 101 & $-12.34 \pm 0.04$ & & 12.98 & 101 & $-12.36 \pm 0.04$ \\
13.1 - 13.3 & 13.19 & 154 & $-12.69 \pm 0.03$ & & 13.19 & 76 & $-12.68 \pm 0.05$ & & 13.19 & 78 & $-12.69 \pm 0.05$ \\
13.3 - 13.5 & 13.38 & 95 & $-13.11 \pm 0.04$ & & 13.38 & 61 & $-12.99 \pm 0.06$ & & 13.38 & 34 & $-13.26 \pm 0.07$ \\
13.5 - 13.7 & 13.58 & 58 & $-13.53 \pm 0.06$ & & 13.58 & 32 & $-13.48 \pm 0.08$ & & 13.57 & 26 & $-13.59 \pm 0.09$ \\
13.7 - 13.9 & 13.78 & 23 & $-14.14 \pm 0.09$ & & 13.74 & 8 & $-14.29 \pm 0.15$ & & 13.82 & 15 & $-14.03 \pm 0.11$ \\
13.9 - 14.1 & 13.96 & 8 & $-14.80 \pm 0.15$ & & 13.97 & 5 & $-14.70 \pm 0.19$ & & 13.95 & 3 & $-14.94 \pm 0.25$ \\
14.1 - 14.4 & 14.17 & 11 & $-15.10 \pm 0.13$ & & 14.17 & 5 & $-15.13 \pm 0.19$ & & 14.22 & 6 & $-15.07 \pm 0.18$ \\
14.4 - 15.05 & 14.64 & 7 & $-16.14 \pm 0.16$ & & 14.71 & 4 & $-16.08 \pm 0.22$ & & 14.43 & 3 & $-16.22 \pm 0.25$ \\
\hline
\end{tabular}
\end{table*}

\begin{table}
\caption{Best-fit models of CDDF for the HIERACHY/MIKE \ion{C}{4} absorption components.}\label{tab:tb_cddf_fit_comp}
\centering
\renewcommand{\arraystretch}{1.3}
\begin{tabular}{ccccc}
\hline
\hline
$z$ range & log$N_{\rm crit}$ & log$f_{13}$ & $\alpha$ & $\beta$ \\ 
\hline
3.0-5.0 &$13.02^{+0.16}_{-0.14}$& $-12.26^{+0.15}_{-0.26}$& $-0.68^{+0.43}_{-0.60}$& $-2.43^{+0.15}_{-0.13}$ \\ \hline
3.0-4.0 &$12.96^{+0.15}_{-0.21}$& $-12.21^{+0.25}_{-0.30}$& $-0.38^{+0.61}_{-0.58}$& $-2.34^{+0.19}_{-0.19}$ \\ \hline
4.0-5.0 &$13.09^{+0.24}_{-0.29}$& $-12.31^{+0.20}_{-0.29}$& $-0.95^{+0.40}_{-0.63}$& $-2.67^{+0.40}_{-0.28}$ \\ \hline
\end{tabular}
\end{table}

\begin{table}
\caption{The $dn_{\rm CIV}/dX$ and $\Omega_{\rm CIV}$ for the HIERACHY/MIKE absorption components in different column density ranges.}\label{tab:dnomegadx_comp}
\centering
\begin{tabular}{cccccc}
\hline\hline
$z_{\rm Med}$ & $z$ range & $\Delta X$ & $\mathscr{N}$ & $dn/dX$ & $\Omega_{\rm CIV} (10^{-8})$  \\ \hline
\multicolumn{6}{c}{$ 12.3 < \log (N/{\rm cm}^2) < 13.2 $} \\ \hline
3.586 &3.000-3.815 & 25.4 & 267 & $8.99\pm0.55$ & $0.92\pm0.06$ \\ \hline
4.034 &3.815-4.215 & 23.3 & 258 & $8.63\pm0.54$ & $0.86\pm0.05$ \\ \hline
4.415 &4.215-5.000 & 33.4 & 247 & $8.66\pm0.56$ & $0.84\pm0.05$ \\ \hline
\multicolumn{6}{c}{$ 13.2 < \log (N/{\rm cm}^2) < 13.8 $} \\ \hline
3.519 &3.000-3.815 & 34.2 & 101 & $2.92\pm0.29$ & $1.17\pm0.12$ \\ \hline
4.019 &3.815-4.215 & 34.1 & 73 & $2.06\pm0.24$ & $0.88\pm0.10$ \\ \hline
4.441 &4.215-5.000 & 35.0 & 62 & $1.80\pm0.23$ & $0.72\pm0.09$ \\ \hline
\multicolumn{6}{c}{$ 13.8 < \log (N/{\rm cm}^2) < 15.2 $} \\ \hline
3.532 &3.000-3.815 & 36.2 & 10 & $0.27\pm0.09$ & $0.92\pm0.29$ \\ \hline
4.023 &3.815-4.215 & 37.5 & 10 & $0.27\pm0.08$ & $1.15\pm0.36$ \\ \hline
4.411 &4.215-5.000 & 35.9 & 15 & $0.42\pm0.11$ & $0.72\pm0.19$ \\ \hline
\multicolumn{6}{c}{$ 13.2 < \log (N/{\rm cm}^2) < 15.0 $} \\ \hline
3.524 &3.000-3.815 & 34.2 & 111 & $3.19\pm0.30$ & $2.09\pm0.20$ \\ \hline
4.020 &3.815-4.215 & 37.5 & 82 & $2.30\pm0.25$ & $1.59\pm0.17$ \\ \hline
4.416 &4.215-5.000 & 35.0 & 77 & $2.22\pm0.25$ & $1.44\pm0.17$ \\ \hline
\end{tabular}
\end{table}

\section{Telluric redshift bin}\label{sec:zbin_tel}

\begin{figure*}
\begin{center}
\includegraphics[width=0.49\textwidth]{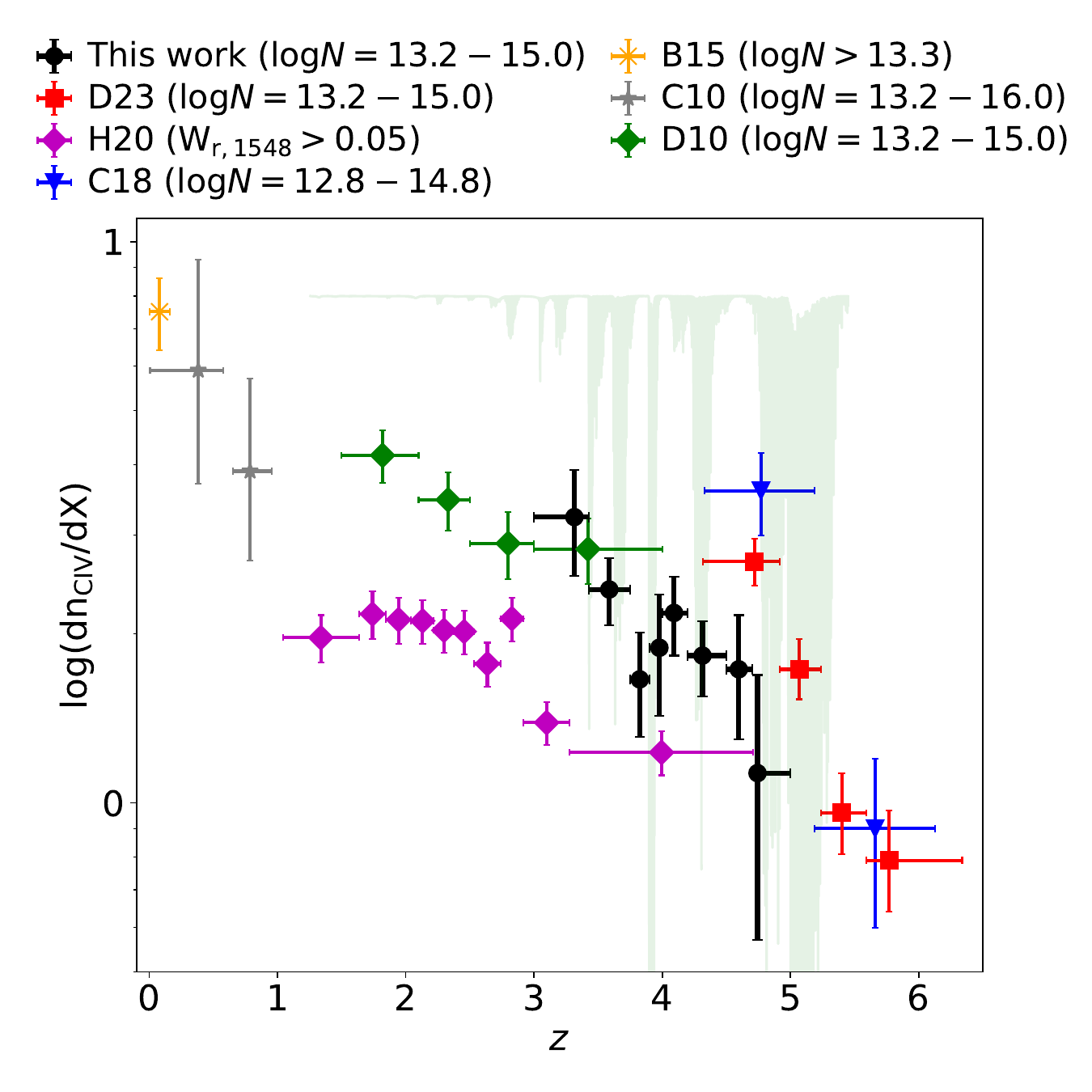}
\includegraphics[width=0.49\textwidth]
{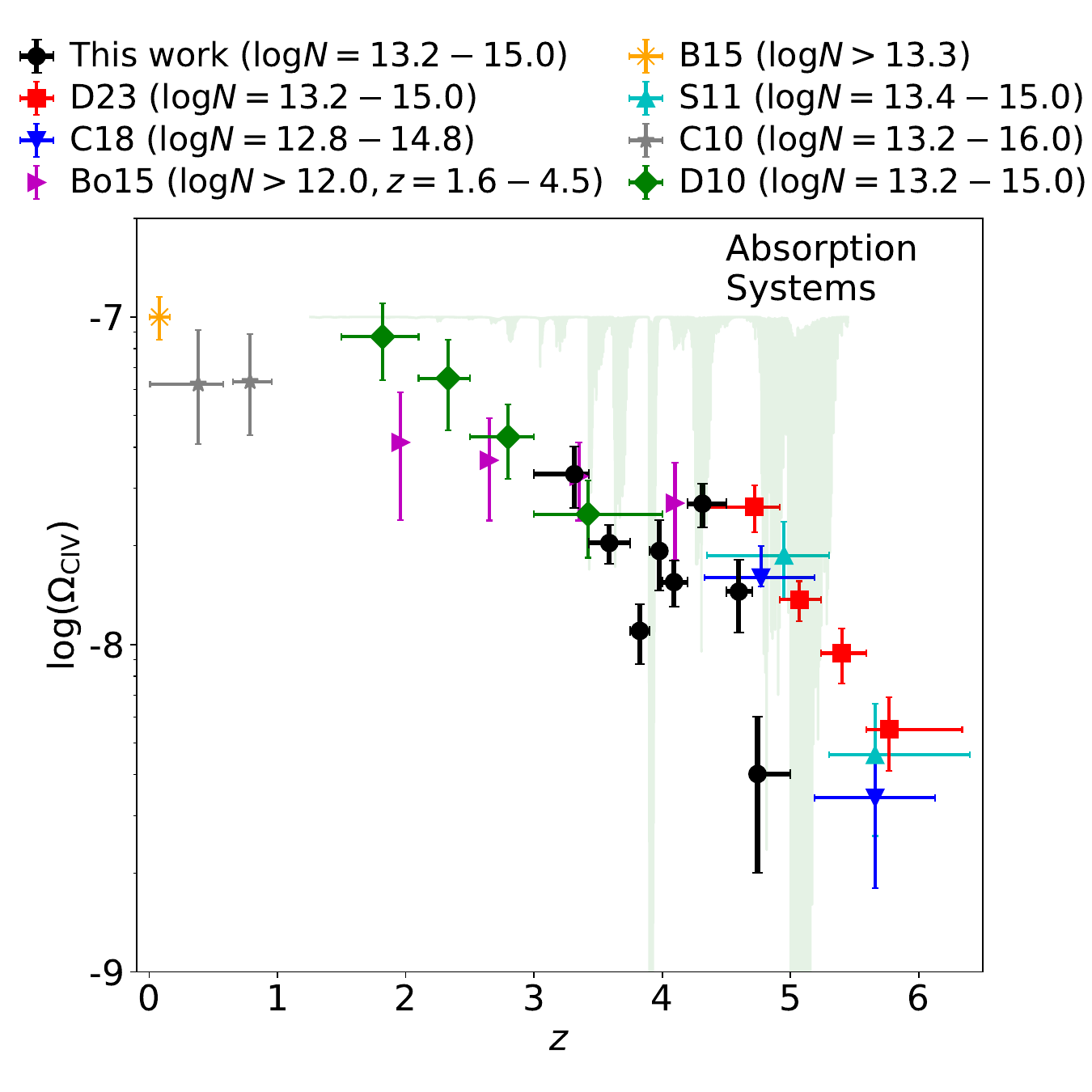}
\caption{The mesured $dn_{\rm CIV}/dX$ (left) and $\Omega_{\rm CIV}$ (right) for the HIERACHY/MIKE \ion{C}{4} absorption systems} with the redshift bin edges determined by telluric contamination (green shaded regions). 
\label{fig:evol_telzbin}
\end{center}
\end{figure*}

The robustness of conclusions on the evolution of $dn_{\rm CIV}/dX$ and $\Omega_{\rm CIV}$ of absorption systems (\S\ref{sec:dndx} and \S\ref{sec:omega}) depends on the goodness of reducing detection bias cased by the widespread and significantly variable telluric contamination.
This telluric contamination could affect the measured \ion{C}{4} absorption line statistics during the \ion{He}{2} EoR (e.g., Fig. \ref{fig:complt}).
Although above measurements contain completeness correction in these contaminated regions, 
we still test the robustness of conclusions in \S\ref{sec:dndx} and \S\ref{sec:omega} via dividing whole samples into eight redshift bins, with bin edges determined by the strength of telluric contamination (Fig. \ref{fig:evol_telzbin}). These measurements support the main conclusions of observed evolution for $dn_{\rm CIV}/dX$ and $\Omega_{\rm CIV}$.
The outlier data points in Fig. \ref{fig:evol_telzbin} are generally from redshift bins with a small bin width or a low number of detected absorbers. 

\end{document}